\documentclass[pre,aps,reprint,amsmath,showkeys]{revtex4-1}

\usepackage[T1]{fontenc} 
\usepackage{graphicx}
\usepackage[utf8]{inputenc}
\usepackage{lmodern}
\usepackage{bm}
\usepackage{bbold}
\usepackage[hidelinks]{hyperref}				
\usepackage[dvipsnames]{xcolor}
\usepackage{ulem}

\newcommand{\rv}{{\bf r}}
\newcommand{\pv}{{\bf p}}
\newcommand{\Fv}{{\bf F}}
\newcommand{\fv}{{\bf f}}
\newcommand{\Jv}{{\bf J}}

\newcommand{\ev}{{\bf e}}

\newcommand{\Tr}{{\rm Tr}}
\newcommand{\eps}{{\boldsymbol\epsilon}}

\newcommand{\tot}{{\rm o}}

\begin{document}

\title{Why Noether's Theorem applies to Statistical Mechanics}

\author{Sophie Hermann}
\email{Sophie.Hermann@uni-bayreuth.de}
\affiliation{Theoretische Physik II, Physikalisches Institut, 
  Universit{\"a}t Bayreuth, D-95447 Bayreuth, Germany\\
  \href{https://www.mschmidt.uni-bayreuth.de/}{\rm www.mschmidt.uni-bayreuth.de}
}

\author{\href{https://www.mschmidt.uni-bayreuth.de/}{Matthias Schmidt}}
\email{Matthias.Schmidt@uni-bayreuth.de}
\affiliation{Theoretische Physik II, Physikalisches Institut, 
  Universit{\"a}t Bayreuth, D-95447 Bayreuth, Germany\\
  \href{https://www.mschmidt.uni-bayreuth.de/}{\rm www.mschmidt.uni-bayreuth.de}
}

\date{7 November 2021; revised version: 11 February 2022}

\begin{abstract}
Noether's Theorem is familiar to most physicists due its fundamental
role in linking the existence of conservation laws to the underlying
symmetries of a physical system. Typically the systems are described
in the particle-based context of classical mechanics or on the basis
of field theory.  We have recently shown
[\href{https://doi.org/10.1038/s42005-021-00669-2}
  {Commun.\ Phys.\ {\bf 4}, 176 (2021)}] that Noether's reasoning also
applies to thermal systems, where fluctuations are paramount and one
aims for a statistical mechanical description.  Here we give a
pedagogical introduction based on the canonical ensemble and apply it
explicitly to ideal sedimentation. The relevant mathematical objects,
such as the free energy, are viewed as functionals. This vantage point
allows for systematic functional differentiation and the resulting
identities express properties of both macroscopic average forces and
molecularly resolved correlations in many-body systems, both in and
out-of-equilibrium, and for active Brownian particles. To provide
further background, we briefly describe the variational principles of
classical density functional theory, of power functional theory, and
of classical mechanics.
\end{abstract}

\maketitle

\section{Introduction}
Symmetries and their breaking in often stunningly beautiful ways are
at the core of a broad range of phenomena in physics, from phase
transitions in condensed matter to mass generation via the Higgs
mechanism. Most readers will be very familiar with the importance of
symmetry operations, including complex operations such as
CPT-invariance in high energy physics as well as the simple challenge
of centering the webcam while having mirroring switched off in a video
call.

The exploitation of the underlying symmetries of a physical system is
an important and central concept that allows to simplify the
mathematical description and arguably more importantly to gain
physical insights and achieve an understanding of the true mechanisms
at play. This is what the mathematician Emmy Noether did in her
groundbreaking work in functional analysis early in the twentieth
century \cite{noether1918}.

Noether analyzed carefully the changes that occur upon performing a
symmetry operation on a system. Her work solved the then open deep
problems of energy conservation in general relativity, as the new
theory of gravity that Einstein had just formed.  Noether considered
the formulation of general relativity via Hilbert's action integral,
which is a formal object --a functional-- that generates Einstein's
field equations.  Nowadays Noether's theorems
\cite{noether1918,neuenschwander2011,byers1998} are widely known and
used to connect each continuous symmetry of a system with a
corresponding conservation law. Noether's work therefore forms a
staple of physics, relevant from introductory classical mechanics to
advanced theories such as the standard model of high energy particle
physics.

In practice the theorems are usually applied to the action functional
in a Lagrangian or Hamiltonian theory. This strategy is not of mere
historic interest, as much active current research is being carried
out, see e.g.\ recent developments that addressed the action
functional for systems that include random forces
\cite{marvian2014quantum,lezcano2018stochastic,baez2013markov} and
work that shows, starting from the symmetry of an action functional,
that the thermodynamic entropy can be viewed as a Noether invariant
\cite{sasa2016,sasa2019,sasa2020}. However, from a mathematical point
of view, Noether's theorem is actually not restricted to the specific
case of the action integral. The theorem rather applies to a much more
general class of functionals, where it specifies general consequences
of invariance under continuous symmetry transformations.

We recall some basics of functional calculus.  A functional is a
mathematical object that maps an entire function, i.e.\ the function
values together with the corresponding values of the argument, to a
single number.  A popular introductory example of a functional is the
definite integral, say over the unit interval from 0 to 1. When viewed
as a functional, the definite integral accepts the integrand (a
function) and it returns a number (the area under the curve that the
function represents).  Although the functional point of view might
appear slightly uncommon (or even trivial in this case), the inherent
abstract concept allows to formulate very significant insights and use
powerful mathematical techniques of variational calculus which can be
straightforwardly and widely applied.

The occurrence of functionals in physics is not restricted to the
study of behaviour at very large length scales, such as that of the
cosmos in the case of general relativity, or to very high energies, as
is the case for fundamental theories of elementary particles.  In fact
the mathematical concept of a functional dependence is very
general. Hence there is an according wide variety of objects in
physics, such as e.g.\ the partition sum and the free energy in
Statistical Mechanics that can be viewed as being a
functional~\cite{mermin1965,evans1979,evans1992,hansen2013}; we give
an introduction below. As soon as one is willing to accept this
notion, making much headway is possible by analyzing physical
properties of the considered system from this formal point of view.

To perform the transfer and use Noether's theorem for thermal systems,
from a formal point of view one would need both to identify a suitable
functional as well as a symmetry transformation under which this
functional is invariant.  One primary candidate for the choice of the
functional is the partition function, which constitutes an integral
over the high-dimensional phase space of classical mechanics. Within
this context, phase space describes all degrees of freedom, i.e.\ the
positions and momenta of all particles in the system. The partition
sum itself is hence an integral over all these variables. Its
integrand is, up to a constant, the Boltzmann factor of the energy
function that characterizes the system. So the partition sum actually
complies with the nature of a functional as it maps this function to
just a number, i.e.\ the value of the partition sum.  (As detailed
below the interesting functional dependence is that on the external
potential.) The partition sum is arguably the most fundamental object
in Statistical Physics, as all thermodynamic quantities, such as
thermodynamic potentials including the free energy, the equation of
state, but also position-resolved correlation functions can be
obtained from it, at least in principle.

Within Statistical Mechanics, where one identifies the free energy
with the negative logarithm of the partition sum, ordinary
(parametric) derivatives of the free energy with respect to
e.g.\ temperature and other thermodynamic variables generate
thermodynamic quantities \cite{evans1979,evans1992,hansen2013}. While
the familiar process of building the derivative of a function, as
giving a measure of the local slope, is a concept that dates back to
Newton and Leibniz, functional differentiation is slightly less
common. However, functionals can be differentiated in much the same
way that functions can be differentiated.  In case of the free energy,
functional derivatives give microscopically resolved correlation
functions \cite{evans1979,evans1992,hansen2013}. These are quantities,
such as the structure factor of a liquid, that are measurable in a
lab, say with a scattering apparatus or even with a microscope upon
further data processing.

When applying Emmy Noether's thinking to the free energy, one could
expect mere abstraction to result, but that is not the case
\cite{hermann2021noether}. Consider the invariance under a spatial
shift.  This classical application of Noether's theorem to the action
functional yields the well-known result of momentum conservation.
When rather exploiting the invariance of the partition function and
hence of the free energy with respect to shifting, what follows are
fundamental statements about forces that act in the system
\cite{hermann2021noether}. One of them states that the total internal
force vanishes. Here the total internal force is that which arises
from the interactions only {\it between} the constituents of the
system. The famous Baron Munchausen tale of bootstrapping himself out
of the swamp by pulling on his own hair is identified as a fairytale
by the Noetherian argument.  The impossibility of this feat holds on
the scale of his entire body, but also when locally resolving his
structure on the molecular scale.

In addition to shifting, one can also consider rotations.  In case of
the action functional being invariant under rotations Noether's
theorem implies that the angular momentum around the rotation axis is
conserved.  If the free energy has rotational symmetry, fundamental
statements about torques emerge
\cite{tarazona1983,hermann2021noether}. These sum rules express
inherent coupling of spin and orbital degrees of freedom.
Figuratively speaking, the identities state that a bolt cannot screw
itself into the wall and that a Baron Munchausen stuck in mud cannot
spontaneously start to rotate by twisting his head.

Recognizing the functional dependence of the free energy allows to
build a theory fully founded on a variational principle of thermal
systems, as formulated by Mermin \cite{mermin1965} and Evans
\cite{evans1979,evans1992}. Their so-called density functional theory
is a well-accepted and widely used theory, see Ref.~\cite{hansen2013}
for a textbook presentation and Ref.~\cite{evans2016specialIssue} for
an overview of recent work. Excellent approximations are available for
relevant model fluids, such as for hard spheres
\cite{rosenfeld1989,roth2010review} (see Ref.~\cite{lin2021} for
recent work addressing hard sphere crystal properties).  The density
functional approach hence allows explicit calculations to be carried
out to predict the behavior of a wide range of physical systems,
including solvation \cite{levesque2012jcp}, hydrophobicity
\cite{jeanmairet2013jcp,evans2019pnas,evans2015prl}, critical drying
of liquids \cite{evans2016prl}, solvent fluctuations
\cite{chacko2017}, electrolyte solutions near surfaces
\cite{martinjimenez2017natCom}, interpretation of atomic force
microscopy data \cite{hernandez-munoz2019}, temperature gradients at
fluid interfaces \cite{muscatello2017}, and local fluctuations
\cite{evans2019pnas,evans2015prl,evans2016prl,chacko2017,
  eckert2020auxiliaryFields}.  In Ref.~\cite{hermann2021noether} we
also apply Noether's thinking to a very recent variational approach
for dynamics, called power functional theory
\cite{schmidt2013pft,schmidt2021pft}, which propels the functional
concepts from equilibrium to nonequilibrium
\cite{schmidt2013pft,schmidt2021pft,
  fortini2014prl,krinninger2019jcp,krinninger2016prl,
  hermann2019prl,hermann2019pre,hermann2020longActive,
  hermann2020polarization,delasheras2018velocityGradient,stuhlmueller2018prl,
  delasheras2020prl,brader2013noz,
  brader2014noz,treffenstaedt2020shear,treffenstaedt2021dtpl},
including the recently popular active Brownian particles
\cite{farage2015, paliwal2018, paliwal2017activeLJ,
  brady2014,soeker2021,auschra2021,hermann2018activeSedimentation}. The
generalization is important, as it shows that not only a dead
Munchausen cannot bootstrap himself out of his misery, but that being
alive does not help (in this particular case).

In the present contribution we demonstrate that the concepts of
Ref.~\cite{hermann2021noether} apply to the canonical ensemble, as is
relevant for confined systems
\cite{gonzalez1997,gonzalez1998,delasheras2014fullCanonical} and for
the dynamics
\cite{delasheras2016particleConservation,schindler2019,wittmann2021}.
Hence having an open system with respect to particle exchange is not
necessary for the Noetherian arguments to apply. We give a detailed
and somewhat pedagogical derivation of the fundamental concepts and
also make much relevant background explicit, which has not been
spelled out in Ref.~\cite{hermann2021noether}.

The paper is organized as follows.  In Sec.~\ref{SECtheory} we go into
some detail and we present in the following the arguably simplest
example of the application of Noether's Theorem to Statistical
Mechanics. We expect the reader to be familiar with Newtonian
mechanics and to ideally know about Classical Mechanics formulated in
a more formal setting (we supply some basic notions thereof below).
We lay out the canonical ensemble and averages in
Sec.~\ref{SECensemble}.  Forces and their relation to symmetries are
addressed in Sec.~\ref{SECforces}. Statistical functionals and their
invariances are described in Sec.~\ref{SECfunctionals}. As an example
we describe the application to sedimentation in
Ref.~\ref{SECapplication}.  The relationship of the Noether invariance
to correlation functions is laid out in
Sec.~\ref{SECrelationshipToCorrelationFunctions}.  We give further
background that is relevant for Ref.~\cite{hermann2021noether}, such
as the details of the grand canonical treatment and the variational
principles of density functional theory and of power functional
theory, in Sec.~\ref{SECdftpft}. We present our conclusions in
Sec.~\ref{SECconclusions}.

\section{Theory}
\label{SECtheory}
\subsection{Canonical ensemble and averages}
\label{SECensemble}
We consider a system with fixed number of particles $N$. The state of
the system is characterized by all positions $\rv_1,\ldots,\rv_N$ and
momenta $\pv_1,\ldots,\pv_N$, where the subscript labels the $N$
particles, which we take to all have identical properties. We assume
that the total energy consists of kinetic and potential energy
contribution, according to
\begin{align}
  H = \sum_{i=1}^N \frac{\pv_i^2}{2m} + u(\rv_1,\ldots,\rv_N) 
  + \sum_{i=1}^N V_{\rm ext}(\rv_i).
  \label{EQhamiltonianNoether}
\end{align}
Here $H$ is the Hamiltonian of the system, with the interparticle
interaction potential $u(\rv_1,\ldots,\rv_N)$ and the external
one-body potential $V_{\rm ext}(\rv_i)$ acting on particle $i$. The
equations of motion are generated via $\dot \rv_i = \partial
H/\partial \pv_i$ and $\dot \pv_i = -\partial H/\partial \rv_i$, where
the overdot indicates a time derivative, $m$ is the particle mass, and
the index $i=1,\ldots,N$. Using the explicit form
\eqref{EQhamiltonianNoether} of the Hamiltonian then leads to the
equations of motion in the familiar form
\begin{align}
  \dot \rv_i &= \frac{\pv_i}{m},
  \label{EQdefinitionVelocityNoether}\\
  \dot \pv_i &= \fv_i,
  \label{EQNewtonTwoNoether}
\end{align}
where $\fv_i$ indicates the force on particle $i$, which consists of a
contribution from all other particles as well as the external
force. Explicitly, the force on particle $i$ is given by
\begin{align}
  \fv_i &= -\nabla_i u(\rv_1,\ldots,\rv_N) 
  - \nabla_i V_{\rm ext}(\rv_i),
  \label{EQforceParticleiNoether}
\end{align}
where $\nabla_i$ denotes the derivative with respect to
$\rv_i$. (Building the derivative by a vector implies building the
derivative with respect to each component of the vector, hence
$\nabla_i$ can be viewed as building the gradient with respect to
$\rv_i$.)  Certainly we could have written down the equations of
motion \eqref{EQdefinitionVelocityNoether} and
\eqref{EQNewtonTwoNoether} a priori. Equation
\eqref{EQdefinitionVelocityNoether} expresses the standard relation of
velocity $\dot \rv_i$ with momentum $\pv_i$, and
\eqref{EQNewtonTwoNoether} is Newton's second law. Hence we have
reproduced the Newtonian theory within the Hamiltonian formalism.

So far everything has been deterministic and we were concerned with
obtaining a description on the level of individual particles. As our
aim is to describe very large systems, we wish to ``zoom out'' and
investigate and describe the macroscopic properties of the system, as
they result from the above formulated microscopic picture. Statistical
Mechanics provides the means for doing so. We will not attempt to give
a comprehensive description of the concepts of this theory. Rather we
will guide the reader through some essential steps, including in
particular how thermal averages are built, to see how Noether's
Theorem applies in this context. As we will see, both the physical
concept and the outcome are different from the standard application of
Noether's theorem based on the action expressed as a time integral
over a Lagrangian that corresponds to \eqref{EQhamiltonianNoether}; we
give a brief description of this standard argument at the end of
Sec.~\ref{SECforces}.

Statistical Mechanics rests on the concept of having a statistical
ensemble, in the sense of the collection of microstates
$\rv_1,\ldots,\rv_N,\pv_1,\ldots,\pv_N$, i.e.\ all phase space
points. These are transcended beyond Classical Mechanics by each being
assigned a probability for its occurrence. (There is much discussion
about who throws the dice here; we recommend Zwanzig's cool-headed
account \cite{zwanzig2001}.) The microstate probability distribution
is given by a standard Boltzmann form,
\begin{align}
  \Psi(\rv_1,\ldots,\rv_N,\pv_1,\ldots,\pv_N)
  &= \frac{{\rm e}^{-\beta H}}{Z_N},
  \label{EQprobabilityDistributionNoether}
\end{align}
where the inverse temperature is $\beta=1/(k_BT)$, with the Boltzmann
constant $k_B$ and absolute temperature~$T$. Here $Z_N$ is the
partition sum, and it acts to normalize the probability distribution
to unity, when summed up over all microstates. The sum over
microstates is in practice a high-dimensional integral over phase
space, explicitly given as
\begin{align}
  Z_N &= \frac{1}{N!h^{3N}} \int d\rv_1\ldots d\rv_N
  d\pv_1\ldots d\pv_N {\rm e}^{-\beta H},
  \label{EQpartitionSumNoether}
\end{align}
where $h$ indicates the Planck constant. Here each position integral
and each momentum integral runs over ${\mathbb R}^3$. (We are
considering systems in three spatial dimensions.) The system volume is
rendered finite by confining walls that are modelled by a suitable
form of the external potential $V_{\rm ext}(\rv)$. As a note on units,
recall that $h$ carries energy multiplied by time, i.e.\ $\rm Js$,
such that the partition sum \eqref{EQpartitionSumNoether} carries no
units.

The purpose of the probability distribution
\eqref{EQprobabilityDistributionNoether} is to build averages. Taking
the Hamiltonian \eqref{EQhamiltonianNoether} as an example, we can
express the total energy, averaged over the statistical ensemble, as
\begin{align}
  E &= \frac{1}{N!h^{3N}}
  \int d\rv_1\ldots\rv_N d\pv_1\ldots d\pv_N H \Psi.
  \label{EQaverageEnergyNoether}
\end{align}
Here we recall the dependence of the Hamiltonian
\eqref{EQhamiltonianNoether} on the phase space point, and in the
notation we have left away the arguments
$\rv_1\ldots,\rv_N,\pv_1,\ldots,\pv_N$ of both $H$ and~$\Psi$.

It is useful to introduce more compact notation, as this reduces
clutter and allows to express the structure of the theory more
clearly. Let us denote the integral over phase space, together with
its normalizing factor in \eqref{EQpartitionSumNoether} as the
``classical trace'' operation, hence defined as
\begin{align}
  \Tr_N  &= \frac{1}{N!h^{3N}}
  \int d\rv_1\ldots d\rv_N
  d\pv_1\ldots d\pv_N,
  \label{EQcanonicalTraceNoether}
\end{align}
which is to be understood as acting on an integrand, such as on
$H\Psi$ in the example \eqref{EQaverageEnergyNoether} above. Equation
\eqref{EQaverageEnergyNoether} can hence be expressed much more
succinctly as
\begin{align}
  E &= \Tr_N H \Psi.
  \label{EQinternalEnergyNoether}
\end{align}
In a similar way we can express other averaged quantities, such as the
average external (potential) energy,
\begin{align}
  U_{\rm ext} &= \Tr_N \Psi \sum_{i=1}^N V_{\rm ext}(\rv_i).
  \label{EQaverageExternalEnergyNoether}
\end{align}
In order to build some trust for the compact notation, we use
\eqref{EQprobabilityDistributionNoether} and
\eqref{EQcanonicalTraceNoether} to re-write
\eqref{EQaverageExternalEnergyNoether} explicitly as
\begin{align}
  U_{\rm ext} &=
  \frac{1}{N!h^{3N}}
  \int d\rv_1\ldots d\rv_N d\pv_1\ldots d\pv_N \notag\\
  &\qquad \qquad\qquad \times
  \frac{{\rm e}^{-\beta H}}{Z_N}
  \sum_{i=1}^N V_{\rm ext}(\rv_i).
  \label{EQaverageExternalEnergyNoetherExplicit}
\end{align}
This allows to see explicitly that $U_{\rm ext}$ depends on the
number of particle $N$ and on the temperature $T$ (via the Boltzmann
factor and the partition sum). Surely
\eqref{EQaverageExternalEnergyNoether} allows to see the physical
content, that of an average being carried out, more clearly than
\eqref{EQaverageExternalEnergyNoetherExplicit} and we will continue to
use the compact notation. (Readers who wish to familiarize themselves
more intimately with these benefits are encouraged to put pen to
scratch paper and re-write the following material in explicit
notation.)

\subsection{Forces and symmetries}
\label{SECforces}
Before continuing with thermal concepts, such as the free energy, we
take a detour from standard paths in Statistical Mechanics, and return
to forces. --After all, it was the microscopically and
particle-resolved forces $\fv_i$ in \eqref{EQforceParticleiNoether}
that formed the starting point for the description of the coupled
system. As an example, let us hence consider the total external force
that acts on the system, in the sense that we sum up the external
force that acts on each individual particle, $-\nabla_i V_{\rm
  ext}(\rv_i)$. This accounting results in $-\sum_i \nabla_i V_{\rm
  ext}(\rv_i)$. Note that this expression still applies per
microstate, or in other words, the total external force varies in
general across phase space. As a cautionary note on terminology, we
use throughout the term ``total'' in the above sense of denoting a
global, macroscopic, extensive quantity. This usage is different from
the also frequent meaning of total referring to the sum of intrinsic
and external contributions.

In order to obtain the macroscopic description we need to trace over
phase space and respect the probability for the occurrence of each
given microstate. Hence the average total external force is given by
\begin{align}
  \Fv_{\rm ext}^\tot &= 
  -\Tr_N \Psi \sum_{i=1}^N \nabla_i V_{\rm ext}(\rv_i).
  \label{EQexternalForceAveragedNoether}
\end{align}
Due to the structure of \eqref{EQexternalForceAveragedNoether},
$\Fv_{\rm ext}^\tot$ depends on the number of particles $N$ (via the
upper limit of the sum and the dimensionality of the phase space
integrals), on temperature $T$ (via the thermal distribution $\Psi$,
cf.\ \eqref{EQprobabilityDistributionNoether} and
\eqref{EQpartitionSumNoether}), and it of course also depends on the
form of the function $V_{\rm ext}(\rv)$. Note that the function
$V_{\rm ext}(\rv)$ appears both explicitly in the gradient in
\eqref{EQexternalForceAveragedNoether} as well as in a more hidden
form in the probability distribution $\Psi$,
cf.~\eqref{EQprobabilityDistributionNoether}
and~\eqref{EQhamiltonianNoether}.

\begin{figure}
  \includegraphics[width=0.9\columnwidth,angle=0]{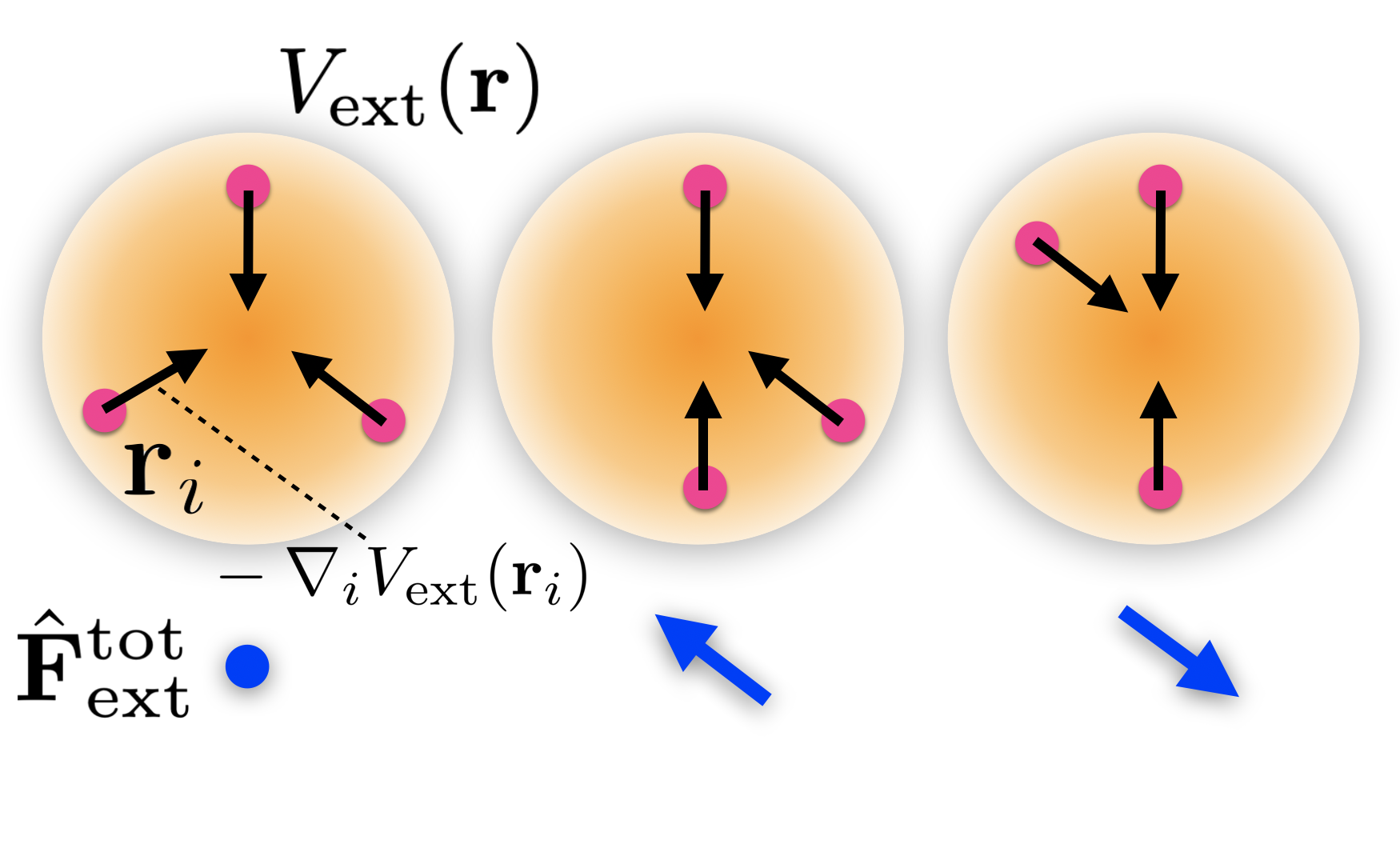}
  \vspace{-0.6cm}
 \caption{Three representative microstates $\rv^N$ for $N=3$ particles
   inside of a spherical cavity modelled by a confining external
   potential $V_{\rm ext}(\rv)$ (orange). Shown are the particle
   positions $\rv_i$ (pink dots) and the respective external force
   $-\nabla_i V_{\rm ext}(\rv_i)$ acting on particle $i$ (black
   arrows). The resulting total external force $\hat\Fv_{\rm
     ext}^\tot=-\sum_i\nabla_i V_{\rm ext}(\rv_i)$ is shown for each
   microstate (blue arrows and blue dot, the later indicating zero).
   Although $\hat\Fv_{\rm ext}^\tot$ for each microstate is in general
   nonzero, the average over the thermal ensemble vanishes, $\Fv_{\rm
     ext}^\tot=\langle \hat\Fv^\tot_{\rm ext}\rangle=0$.  }
  \label{FIG1}
\end{figure}

Let us halt for a moment and ponder the physics. Imagine having a
vessel with impenetrable walls, such that the system stays confined
inside of the vessel. Furthermore, to add some flavour, imagine an
external field such as gravity acting on the system. Then the external
potential consists of two contributions, i.e.\ the potential energy
that the container walls exert on each given particle plus the
gravitational energy. In an equilibrium situation, what would we
expect the total external force to be like?  Surely, it should not
change in time. (Technically any time evolution had been superseded by
the ensemble, which is a static one in the present case.) The reader
might expect that
\begin{align}
  \Fv_{\rm ext}^\tot&=0,
  \label{EQforceZeroNoether}
\end{align}
because otherwise the system would surely start to move!  However, as
for any given microstate the total external force will in general {\it
  not} vanish, \eqref{EQforceZeroNoether}, if true, is a nontrivial
property of thermal equilibrium.  See Fig.~\ref{FIG1} for an
illustration this concept, based on a system confined in a spherical
cavity.  Hence we wish to address carefully in the following whether
we can prove \eqref{EQforceZeroNoether} from first principles.

In the following we give two derivations of
\eqref{EQforceZeroNoether}, which both rest on spatial translations of
the system. The first derivation only requires vector calculus. The
second derivation shows the Noetherian symmetry argument based on the
functional setting. This requires to adopt the notion of functional
dependencies, which we have used only implicitly so far. In the
following we make these relationships and dependencies explicit. We
also supply the necessary methodology of functional differentiation
and will attempt to convince the reader that their background in
ordinary calculus can be flexed in order to follows these steps.

The fundamental ingredients to both derivations are identical
though. We use the free energy and we monitor its changes upon spatial
displacement of the system. The free energy, and more generally
thermodynamic potentials, are central to thermal physics, and the
following material can be viewed as a demonstration why this indeed is
the case.

The free energy $F_N$, or more precisely: the total Helmholtz free
energy is given by
\begin{align}
  F_N &= -k_BT \ln Z_N,
  \label{EQfreeEnergyDefinitionNoether}
\end{align}
where $Z_N$ is the partition sum, as defined in
\eqref{EQpartitionSumNoether}. One can show that the relation of free
energy and internal energy is given by the thermodynamic identity
$F_N=E-TS$, where $S$ is the entropy, here defined on a microscopic
basis and the internal energy~$E$ is given by
\eqref{EQinternalEnergyNoether}. One can surely be surprised by the
promotion of the rather banal normalization factor $Z_N$ to such a
prominent and as we show decisive role. We demonstrate in the
following that $Z_N$ had been a dark horse, and that its status to
generate the free energy via \eqref{EQfreeEnergyDefinitionNoether} is
well-deserved.

Besides the free energy, the second ingredient that we require is a
spatial shift of the entire system according to a displacement vector
$\eps$ of the system. We hence displace the external potential
spatially by a constant vector~$\eps$.  (Although we Taylor expand in
$\eps$ below, the displacement~$\eps$ can be finite and arbitrary.)
The displaced system is then under the influence of an external
potential which has changed according to
\begin{align}
  V_{\rm ext}(\rv) \to V_{\rm ext}(\rv+\eps).
  \label{EQexternalPotentialDisplacementNoether}
\end{align}
Formally, the free energy of the displaced system will depend on the
displacement vector, i.e.\
\begin{align}
  F_N \to F_N(\eps),
\end{align}
where $F_N$ is the free energy \eqref{EQfreeEnergyDefinitionNoether}
expressed in the original coordinates, and the new free energy is
given by
\begin{align}
  F_N(\eps) &= -k_BT\ln Z_N(\eps).
  \label{EQfreeEnergyEpsilonNoether}
\end{align}
Here the partition sum of the shifted system is
\begin{align}
  Z_N(\eps) &=\Tr_N 
  \exp\Big[
  -\beta \Big(H_{\rm int} + 
  \sum_i V_{\rm ext}(\rv_i+\eps)
  \Big)\Big],
  \label{EQpartitionSumEpsilonNoether}
\end{align}
where the intrinsic part $H_{\rm int}$ of the Hamiltonian consists of
kinetic energy and interparticle interaction potential energy only,
i.e.\ $H_{\rm int}=\sum_i \pv_i^2/(2m)+ u(\rv_1,\ldots,\rv_N)$.

We proceed by first recognizing that the shift does not change the
value of the free energy (in other words, the choice of origin of the
coordinate system does not matter). We can see this explicitly by
performing a coordinate transformation $\rv_i \to \rv_i-\eps$. This
leaves $H_{\rm int}$ invariant, as the momenta are unaffected and the
internal interaction potential is unaffected. --Recall that the
interparticle energy only depends on relative particle positions,
which remain invariant under the transformation:
$\rv_i-\rv_j\to(\rv_i-\eps)-(\rv_j-\eps)=\rv_i-\rv_j$. Furthermore,
due to the simplicity of the coordinate transformation that the shift
represents, the phase space integral, cf.\ the classical trace
\eqref{EQcanonicalTraceNoether}, is unaffected as the Jacobian of the
transformation is unity.  Note that in the shifting operation, the
momenta are unaffected and their behaviour remains governed by the
Maxwell distribution throughout.  Hence we have shown that the
original free energy is identical to the free energy of the shifted
system
\begin{align}
  F_N = F_N(\eps),
  \label{EQepsilonInvarianceNoether}
\end{align}
for any value of the displacement vector $\eps$.

At this point one could conclude mission accomplished. This is not
what Emmy Noether did in her mathematical formulation of the problem
-- we hint at her variational techniques below. The way forward at
this point is to rather ignore \eqref{EQepsilonInvarianceNoether} and
return to the explicit expression \eqref{EQfreeEnergyEpsilonNoether}
for the free energy in the shifted system. We consider small
displacements $\eps$ and Taylor expand to first order,
\begin{align}
  F_N(\eps) = F_N + \frac{\partial F_N(\eps)}{\partial \eps}
  \Big|_{\eps=0}\cdot \eps,
  \label{EQTaylorExpansionFNoether}
\end{align}
where quadratic and higher order terms in $\eps$ have been omitted.
The partial derivative in \eqref{EQTaylorExpansionFNoether} can be
calculated explicitly:
\begin{align}
  \frac{\partial F_N(\eps)}{\partial \eps} &=
  -\frac{k_BT}{Z_N(\eps)}
  \frac{\partial}{\partial\eps}Z_N(\eps)
  \label{EQFNderivative1Noether}\\
  &= -\frac{k_BT}{Z_N(\eps)}
  \Tr_N
  \frac{\partial}{\partial\eps}
  {\rm e}^{ -\beta H(\eps)}
  \label{EQFNderivative2Noether}\\
  &= -\frac{k_BT}{Z_N(\eps)}
  \Tr_N
     {\rm e}^{ -\beta H(\eps)}
       \frac{\partial}{\partial\eps}(-\beta)\sum_{i=1}^N 
       V_{\rm ext}(\rv_i-\eps),
       \label{EQFNderivative3Noether}
\end{align}
where in the first step \eqref{EQFNderivative1Noether} the partition
sum in the denominator arises from the derivative of the logarithm in
\eqref{EQfreeEnergyEpsilonNoether} and in the second step
\eqref{EQFNderivative2Noether} we have interchanged the phase space
integration [as notated by $\Tr_N$,
  cf.~\eqref{EQcanonicalTraceNoether}] and the $\eps$-derivative.  The
third step \eqref{EQFNderivative3Noether} follows directly from the
structure of the Hamiltonian \eqref{EQhamiltonianNoether} and the fact
that $H_{\rm int}$ is independent of $\eps$. We continue to obtain
\begin{align}
  \frac{\partial F_N(\eps)}{\partial \eps} &=
  \Tr_N
     \frac{{\rm e}^{ -\beta H(\eps)}}{Z_N(\eps)}
       \sum_{i=1}^N \frac{\partial}{\partial\eps}
       V_{\rm ext}(\rv_i-\eps)
       \label{EQFNderivative4Noether}\\
       &=
       -\Tr_N
       \Psi(\eps)
       \sum_{i=1}^N \frac{\partial}{\partial\rv_i}
       V_{\rm ext}(\rv_i-\eps),
       \label{EQFNderivative5Noether}
\end{align}
where in \eqref{EQFNderivative4Noether} we have pulled the partition
sum as a constant inside of phase space integral and have moved the
$\eps$-derivative inside the sum over all particles.  In
\eqref{EQFNderivative5Noether} we have combined the Boltzmann factor
with the partition sum in order to express the many-body probability
distribution function in the shifted system, $\Psi(\eps)=\exp(-\beta
H(\eps))/Z_N(\eps)$, in generalization of
\eqref{EQprobabilityDistributionNoether}. Furthermore the spatial
derivative of the external potential is re-written via using
$\partial/\partial\eps=-\partial/\partial \rv_i$ (which is valid due
to the dependence on only the difference $\rv_i-\eps$). Considering
the case $\eps=0$ allows us to conclude that
\begin{align}
  \frac{\partial F_N(\eps)}{\partial \eps} \Big|_{\eps=0}&=
       -\Tr_N
       \Psi
       \sum_{i=1}^N \frac{\partial}{\partial\rv_i}
       V_{\rm ext}(\rv_i).
\end{align}
Remarkably the right hand side is the average total external force as
previously defined in \eqref{EQexternalForceAveragedNoether}. The left
hand side is identically zero, as $\eps$ is arbitrary in
\eqref{EQTaylorExpansionFNoether} and the linear order (as well as all
higher orders) need to vanish in the Taylor expansion
\eqref{EQTaylorExpansionFNoether} by virtue of the invariance
\eqref{EQepsilonInvarianceNoether} of the free energy upon spatial
displacement. Hence
\begin{align}
       -\Tr_N
       \Psi
       \sum_{i=1}^N \frac{\partial}{\partial\rv_i}
       V_{\rm ext}(\rv_i) &= 0,
       \label{EQresultOfProofNoether}
\end{align}
which proves constructively the anticipated vanishing
\eqref{EQforceZeroNoether} of the average total external force
\eqref{EQexternalForceAveragedNoether}.

As a preliminary summary, we have shown that the invariance of a
global thermodynamic potential, the Helmholtz free energy expressed in
the canonical ensemble, against spatial displacement (as generated by
a shift of the external potential) leads to the non-trivial force
identity of vanishing total external force. This identity holds true
for any value of the number of particles in the system, at arbitrary
temperature, and most notably irrespective of the precise form of the
external potential. Hence we refer to statements such as $\Fv_{\rm
  ext}^\tot=0$, cf.~\eqref{EQforceZeroNoether}, as a Noether identity
or Noether sum rule. Clearly the concept is general, as both the
symmetry operation can be altered (rotations are considered in
Ref.~\cite{hermann2021noether}) as well as the type of thermodynamic
object can be changed (the grand potential and the excess free energy
density functional are considered in Ref.~\cite{hermann2021noether}
and we shift the total external energy $U_{\rm ext}$ below in
Sec.~\ref{SECrelationshipToCorrelationFunctions}).

We have presented here the shifting from the point of view that the
actual physical system is moved to a different
location. Alternatively, one could adopt a ``passive'' point of view
and displace only the origin of the coordinate system, in the sense of
using shifted coordinates that still describe an unchanged physical
system. Then going through a chain of arguments analog to those given
above yields identical results.

For completeness we contrast the present statistical mechnical
treatment with the standard application of Noether's theorem to
deterministic dynamics. We keep the same $N$-body classical many-body
system as before, i.e.\ with Hamiltonian $H$ given by
\eqref{EQhamiltonianNoether}. The equations of motion
\eqref{EQdefinitionVelocityNoether} and \eqref{EQNewtonTwoNoether}
follow from the action integral ${\cal S}=\int_{t_1}^{t_2}dt L$, where
the Lagrangian $L$ is obtained via $L=\sum_i\pv_i\dot\rv_i-H$ and
$t_1$ and $t_2$ are two fixed points in time. We apply the global
shifting transformation $\rv_i\to\rv_i-\eps$, as before, to all
particle coordinates in the system and at all times. As a consequence,
the Lagrangian acquires a corresponing dependence on $\eps$. Taylor
expanding the action to first order in $\eps$ then yields
\begin{align}
  {\cal S}(\eps) &= {\cal S} + 
  \frac{\partial {\cal S}(\eps)}{\partial \eps}
  \Big|_{\eps=0}\cdot \eps
  \label{EQnoetherAction1}\\
  &= {\cal S} + \int_{t_1}^{t_2} dt \frac{\partial L}{\partial \eps}
  \Big|_{\eps=0}\cdot\eps
  \label{EQnoetherAction2}\\
  &= {\cal S} - \int_{t_1}^{t_2} dt \sum_i\frac{\partial L}{\partial \rv_i}
  \Big|_{\eps=0}
  \cdot\eps
  \label{EQnoetherAction3}\\
  &= {\cal S} - \int_{t_1}^{t_2} dt \sum_i\frac{d\pv_i}{dt}
  \cdot\eps
  \label{EQnoetherAction4}\\
  &= {\cal S} - \sum_i\pv_i \Big|_{t_1}^{t_2}
  \cdot\eps,
  \label{EQnoetherActionFinal}
\end{align}
where ${\cal S}={\cal S}(\eps=0)$ is the action in the original
unshifted system; we have used the representation of ${\cal S}(\eps)$
as the time integral of the Lagrangian in the derivation of
\eqref{EQnoetherAction2}, the identity $\partial
L/\partial\eps=-\sum_i \partial L/\partial \rv_i$ to obtain
\eqref{EQnoetherAction3}, the Lagrangian equations of motion
$d\pv_i/dt=\partial L/\partial\rv_i$ to derive
\eqref{EQnoetherAction4}, and the fact that the integrand of
\eqref{EQnoetherAction4} is a total time differential to obtain
\eqref{EQnoetherActionFinal}.

Suppose now that the system is invariant under the displacement, such
that ${\cal S}={\cal S}(\eps)$ for any value of $\eps$ and the second
term in \eqref{EQnoetherActionFinal} needs to vanish. This implies
that the global momentum ${\bf P}^\tot=\sum_i \pv_i$ is conserved,
i.e.\ ${\bf P}^\tot(t_2)={\bf P}^\tot(t_1)$.

\subsection{Functionals and invariances}
\label{SECfunctionals}
The abstraction that is yet to be performed and that allows to see the
above statistical mechanical force result in an even wider setting, is
based on functional methods. As we had hinted at in the introduction,
integrals often allow for direct interpretation as functionals as they
map their integrand (or part thereof) to the value of the
quadrature. In the specific case at hand, we stay with the canonical
free energy $F_N$ and observe that its value certainly depends on the
form of the external potential $V_{\rm ext}(\rv)$, cf.\ its occurrence
in the Hamiltonian \eqref{EQhamiltonianNoether}, which via the
partition sum \eqref{EQpartitionSumNoether} enters the free energy
\eqref{EQfreeEnergyDefinitionNoether}. Hence we have $F_N[V_{\rm
    ext}]$, where we indicate the functional dependence by square
brackets (and leave away in the notation the position argument $\rv$,
despite the fact that the functional depends on the entire
function). In order to highlight this point of view, we rewrite
\eqref{EQfreeEnergyDefinitionNoether} and
\eqref{EQpartitionSumNoether}, respectively, in the form
\begin{align}
  F_N[V_{\rm ext}] &= -k_BT \ln Z_N[V_{\rm ext}],
  \label{EQfreeEnergyAsFunctionalCompactNoether}\\
  Z_N[V_{\rm ext}] &= 
  \Tr_N \exp\Big(
  -\beta H_{\rm int} -\beta \sum_{i=1}^N V_{\rm ext}(\rv_i)
  \Big),
  \label{EQpartitionSumAsFunctionalNoether}
\end{align}
where still the partition sum, viewed now as a functional of the
external potential, $Z_N[V_{\rm ext}]$ is given by its elementary
form, i.e.\ the right hand side of \eqref{EQpartitionSumNoether}.  In
a more compact form, eliminating $Z_N[V_{\rm ext}]$ as a standalone
object, we have
\begin{align}
  & F_N[V_{\rm ext}] =
  \label{EQfreeEnergyAsFunctionalNoether}\\
  &\qquad
  -k_BT \ln 
  \Tr_N \exp\Big(
  -\beta H_{\rm int} -\beta \sum_{i=1}^N V_{\rm ext}(\rv_i)
  \Big).\notag
\end{align}
We dwell on the functional concept and demonstrate some practical
consequences. As an analogy, viewing the functional dependence in
\eqref{EQfreeEnergyAsFunctionalNoether} akin to the dependence of an
ordinary function $f(x)$ on its argument $x$ brings concepts of
calculus immediately to mind, such as building the derivative $f'(x)$
and investigating its properties.

This analogy extends to functionals and their derivatives with respect
to the argument function, in a process referred to as functional
differentiation. For the present case, functionally deriving $F_{\rm
  ext}[V_{\rm ext}]$ with respect to $V_{\rm ext}(\rv)$ can be viewed
as monitoring the change of the value of the functional upon changing
its argument function at position $\rv$. The change will in general
depend on position $\rv$, hence building functional derivatives
creates position dependence. (The result of the functional derivative
is again a functional, as the dependence on the argument function
persists.)  Functional calculus is in many ways similar to ordinary
multi-variable calculus. We do not attempt to give a tutorial here
(see e.g.\ the appendix of Ref.~\cite{schmidt2021pft} for a very brief
one), but rather present a single example that is relevant for the
present physics of invariance operations applied to many-body systems.

We use standard notation and denote the functional derivative with
respect to the function $V_{\rm ext}(\rv)$ as $\delta/\delta V_{\rm
  ext}(\rv)$. Applying this procedure to the free energy
\eqref{EQfreeEnergyAsFunctionalCompactNoether} yields
\begin{align}
  \frac{\delta F_N[V_{\rm ext}]}{\delta V_{\rm ext}(\rv)} &=
  -k_BT \frac{\delta}{\delta V_{\rm ext}(\rv)} \ln Z_N[V_{\rm ext}]
  \label{EQfunctionalDerivative1Noether}\\
  &= -\frac{k_BT}{Z_N[V_{\rm ext}]}
  \frac{\delta}{\delta V_{\rm ext}(\rv)} Z_N[V_{\rm ext}],
  \label{EQfunctionalDerivative2Noether}
\end{align}
where in the first step we have taken the multiplicative constant
$-k_BT$ out of the derivative and in the second step have used the
ordinary chain rule, which also holds for functional
differentiation. We next use the explicit form
\eqref{EQpartitionSumAsFunctionalNoether} to obtain
\begin{align}
  \frac{\delta F_N[V_{\rm ext}]}{\delta V_{\rm ext}(\rv)} &=
  -\frac{k_BT}{Z_N[V_{\rm ext}]}
  \Tr_N \frac{\delta}{\delta V_{\rm ext}(\rv)}
     {\rm e}^{-\beta H}
     \label{EQfunctionalDerivative3Noether}\\
     &= -\frac{k_BT}{Z_N[V_{\rm ext}]}
     \Tr_N {\rm e}^{-\beta H} \frac{\delta}{\delta V_{\rm ext}(\rv)}
     (-\beta H)
     \label{EQfunctionalDerivative4Noether}\\
     &= \frac{1}{Z_N[V_{\rm ext}]}
     \Tr_N {\rm e}^{-\beta H} \frac{\delta}{\delta V_{\rm ext}(\rv)}
     \sum_{i=1}^N V_{\rm ext}(\rv_i),
     \label{EQfunctionalDerivative5Noether}
\end{align}
where we have first exchanged the order of the functional derivative
and the phase space integral, i.e.\ moved the derivative inside of the
trace in \eqref{EQfunctionalDerivative3Noether}, then in the second
step \eqref{EQfunctionalDerivative4Noether} have used the chain rule
to differentiate the exponential, and in the last step
\eqref{EQfunctionalDerivative5Noether} have exploited the structure
\eqref{EQhamiltonianNoether} of the Hamiltonian. Moving the derivative
inside of the sum over $i$ and identifying the many-body probability
distribution function $\Psi$ according to
\eqref{EQprobabilityDistributionNoether} yields the final result
\begin{align}
  \frac{\delta F_N[V_{\rm ext}]}{\delta V_{\rm ext}(\rv)} &=
  \Tr_N \Psi \sum_i \delta(\rv-\rv_i)
  \equiv \rho(\rv),
  \label{EQdensityProfileNoether}
\end{align}
where we have used one central rule of functional differentiation:
differentiating a function by itself gives $\delta V_{\rm
  ext}(\rv_i)/\delta V_{\rm ext}(\rv)=\delta(\rv-\rv_i)$, where the
result $\delta(\cdot)$ is the Dirac delta distribution (here in three
dimensions, as its argument is a three-dimensional vector).

Notably in \eqref{EQdensityProfileNoether} we have arrived at the form
of a thermal average over the statistical ensemble; recall the generic
form exemplified by the average internal
energy~\eqref{EQinternalEnergyNoether}. Rather than the expectation
value of the Hamiltonian, the present case represents the average of
the microscopically resolved density operator $\sum_{i=1}^N
\delta(\rv-\rv_i)$, which can be viewed as an indicator function that
measures whether any particle resides at the given position $\rv$. The
result of the average is the one-body density distribution, or in
short the density profile $\rho(\rv)$. That functional differentiation
yields useful, spatially-resolved (``correlation'') functions is a
general mechanism. See e.g.\ \cite{hansen2013} for much background on
correlation functions and their generation via functional
differentiation. Reference~\cite{hermann2021noether} carries this
concept much further than we do here.

\begin{figure}
  \includegraphics[width=0.98\columnwidth,angle=0]{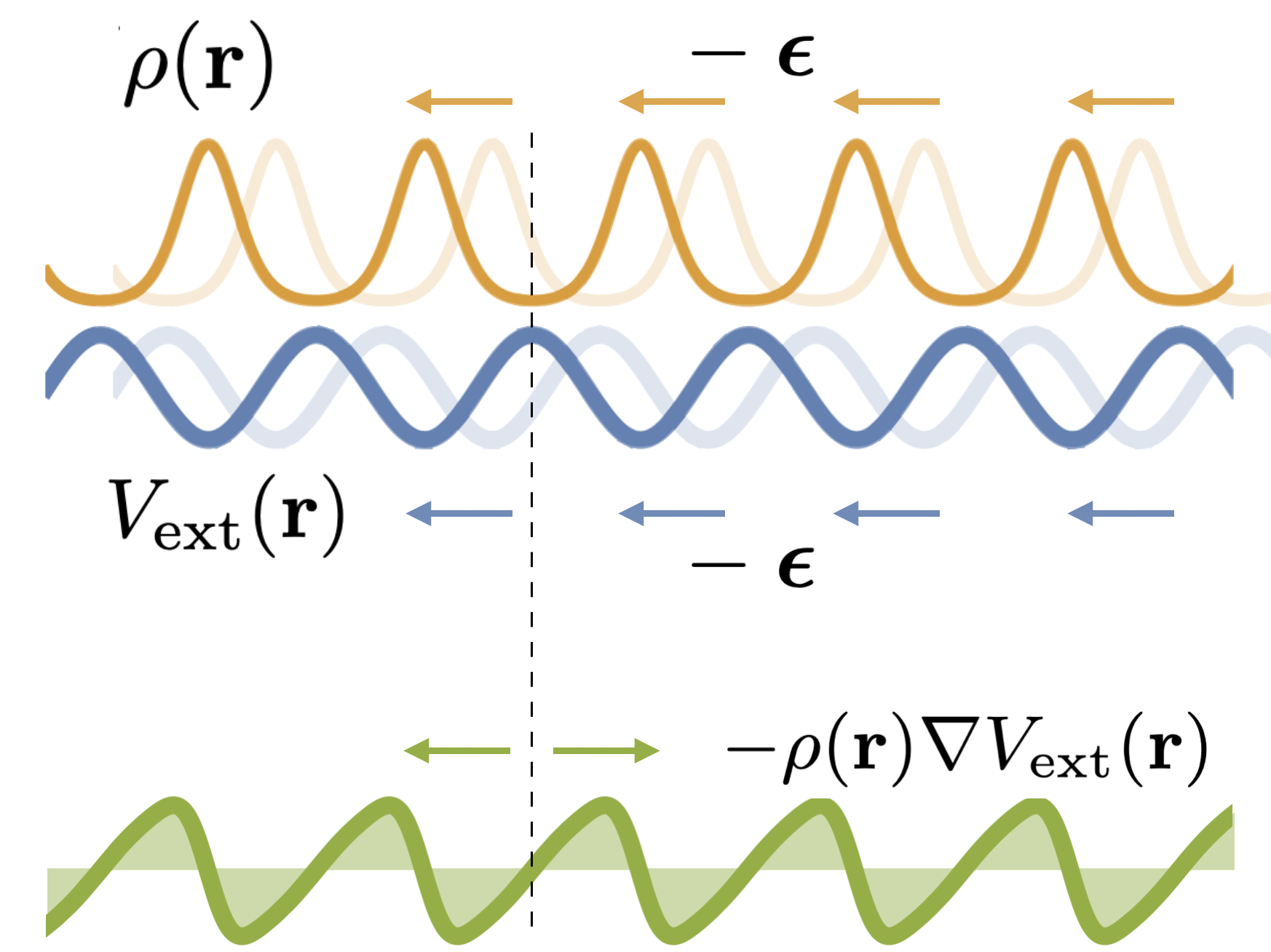}
  \caption{Illustration of the shifting. A sinusoidal external
    potential $V_{\rm ext}(\rv)$ (blue lines) is spatially displaced
    by a displacement $-\eps$ (blue arrows). The density profile
    $\rho(\rv)$ (amber lines) measures the local probability to find a
    particle; it hence has e.g.\ peaks at the troughs of the external
    potential and it is shifted accordingly (amber arrows).  Also shown
    is the magnitude of the external force density $-\rho(\rv)\nabla
    V_{\rm ext}(\rv)$ (green line); the green arrows represent its
    local direction. The horizontal dashed line is a guide that
    indicates the position of locally vanishing external force
    density.}
  \label{FIG2}
\end{figure}

We return to the shifting symmetry operation of above, but now monitor
the system response via tracking the changes in the function $V_{\rm
  ext}(\rv)$ that are induced by the spatial shifting. Recall the
elementary Taylor expansion
\begin{align}
  V_{\rm ext}(\rv+\eps) &= V_{\rm ext}(\rv)
  +\eps\cdot \nabla V_{\rm ext}(\rv),
  \label{EQVextTaylorNoether}
\end{align}
where $\nabla$ indicates the derivative (gradient) with respect to
$\rv$ and we have truncated at linear order.  See Fig.~\ref{FIG2} for
an illustration. The first order term in \eqref{EQVextTaylorNoether}
can be viewed as a local change in the external potential, $\delta
V_{\rm ext}(\rv)$, which is given by
\begin{align}
  \delta V_{\rm ext}(\rv) \equiv \eps\cdot \nabla V_{\rm ext}(\rv).
  \label{EQdeltaVextNoether}
\end{align}

In order to capture the resulting effect on the functional, we can
functionally Taylor expand the dependence of the free energy on
$V_{\rm ext}(\rv)+\delta V_{\rm ext}(\rv)$ around the function $V_{\rm
  ext}(\rv)$.  To linear order in $\delta V_{\rm ext}(\rv)$ the
functional Taylor expansion reads
\begin{align}
  F_N[V_{\rm ext}+\delta V_{\rm ext}]
  &=F_N[V_{\rm ext}] +
  \int d\rv
  \frac{\delta F_N[V_{\rm ext}]}{\delta V_{\rm ext}(\rv)}
  \delta V_{\rm ext}(\rv)
  \label{EQfreeEnergyExpansionPreliminaryNoether}
  \\ &
  =F_N[V_{\rm ext}] +
  \int d\rv
  \rho(\rv)
  \eps\cdot \nabla V_{\rm ext}(\rv),
  \label{EQfreeEnergyExpansionNoether}
\end{align}
where in \eqref{EQfreeEnergyExpansionNoether} we have used the
explicit form \eqref{EQdeltaVextNoether} of $\delta V_{\rm ext}(\rv)$
as it arises from the fact that the variation in the shape of the
external potential is specifically generated by a spatial
displacement, cf.~\eqref{EQVextTaylorNoether}. Furthermore we have
used \eqref{EQdensityProfileNoether} to identify the functional
derivative in \eqref{EQfreeEnergyExpansionPreliminaryNoether} as the
density profile.

The result \eqref{EQfreeEnergyExpansionNoether} is based on the
properties of functional calculus alone. Hence the identity is general
and holds, to linear order in $\eps$, irrespective of any invariance
properties.  For the case of the total free energy, which as we have
shown above in \eqref{EQepsilonInvarianceNoether} is invariant under
spatial displacement, we have
\begin{align}
  F_N[V_{\rm ext}] &= F_N[V_{\rm ext}+\delta V_{\rm ext}],
  \label{EQfreeEnergyInvariantNoether}
\end{align}
where $\delta V_{\rm ext}(\rv)$ is generated from the spatial
displacement of the system, cf.~\eqref{EQdeltaVextNoether}. Hence
\eqref{EQdeltaVextNoether} together with
\eqref{EQfreeEnergyInvariantNoether} express in functional language
the translational symmetry properties of the free energy.

From the identity \eqref{EQfreeEnergyInvariantNoether} and the linear
Taylor expansion \eqref{EQfreeEnergyExpansionNoether} we can conclude
that the correction term needs to vanish,
\begin{align}
  \int d\rv \rho(\rv) \eps\cdot \nabla V_{\rm ext}(\rv) &= 0.
\end{align}
The displacement vector $\eps$ is arbitrary, as there was no
restriction on the direction of the shift. Hence the above expression
can only identically vanish provided that
\cite{baus1984,hermann2021noether}
\begin{align}
  \Fv_{\rm ext}^\tot =
  -\int d\rv \rho(\rv) \nabla V_{\rm ext}(\rv)  &= 0,
  \label{EQFextTotViaDensityProfileNoether}
\end{align}
where we have multiplied by $-1$ in order to identify the one-body
expression for the total external force $\Fv_{\rm ext}^\tot$; the
equivalence with the many-body form
\eqref{EQexternalForceAveragedNoether} is straightforward to show upon
using the definition of the density profile
\eqref{EQdensityProfileNoether}.  See Fig.~\ref{FIG2} for an
illustration of the local force density profile, i.e.\ the integrand
of \eqref{EQFextTotViaDensityProfileNoether}.

\subsection{Application to sedimentation}
\label{SECapplication}
\begin{figure}
  \includegraphics[width=0.9\columnwidth,angle=0]{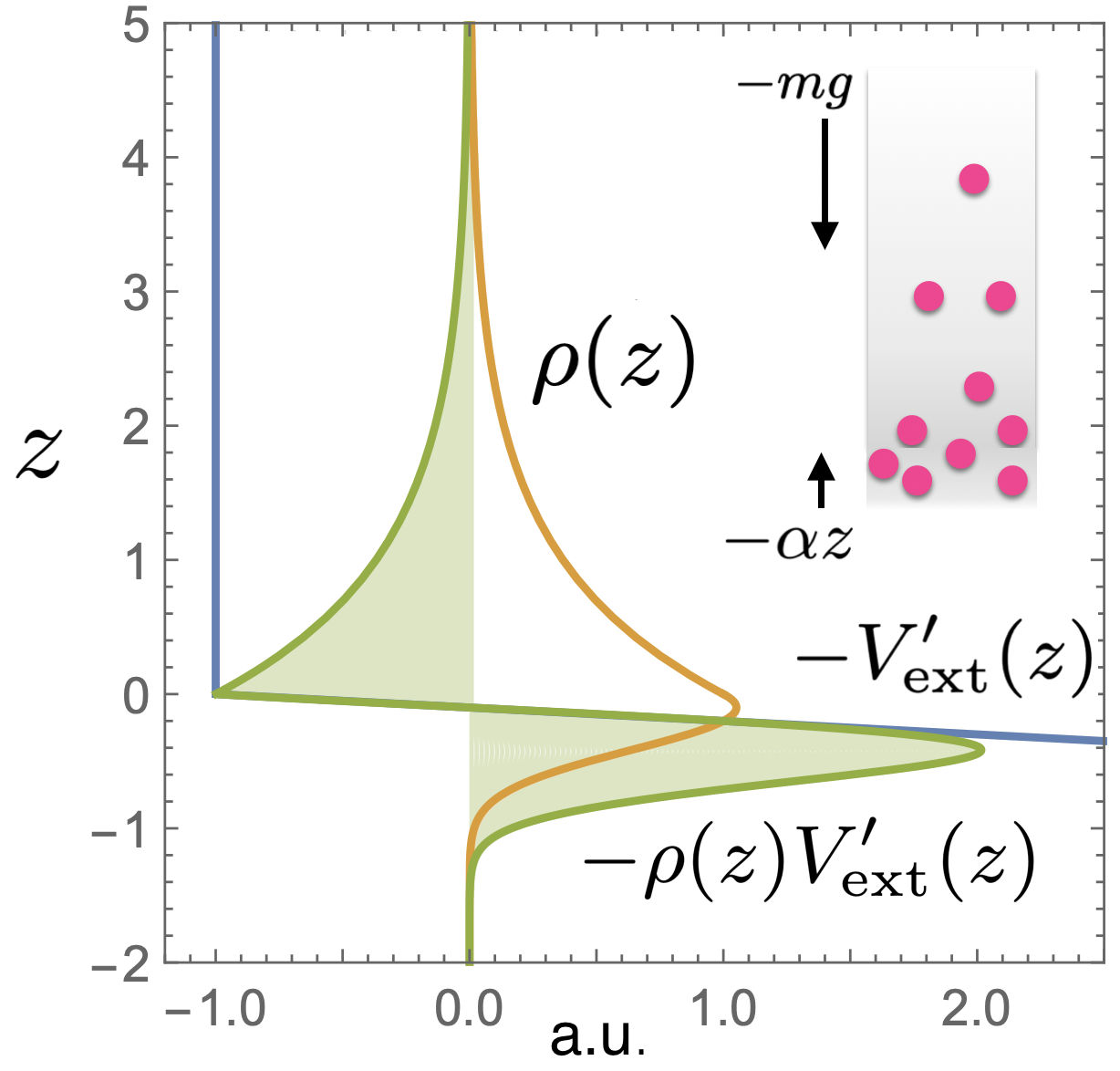}
  \caption{Illustration of sedimentation of a fluid against a lower
    soft wall represented by a harmonic potential.  The total external
    potential is $V_{\rm ext}(z)=mgz + \Theta(-z)\alpha z^2/2$. The
    resulting external force field is $-V_{\rm ext}'(z)\equiv-\partial
    V_{\rm ext}(z)/\partial z = -mg-\Theta(-z)\alpha z$ (blue
    line). The direction of both force contributions is indicated in
    the inset (arrows), where pink dots represent particles. In the
    main plot the density profile $\rho(z)$ (amber line) decays for
    large and for small values of $z$.  The external force density is
    the product $-\rho(z)V_{\rm ext}'(z)$ (green line). The total
    external force (per unit area) is the integral $-\int dz V_{\rm
      ext}'(z)\rho(z)=0$; note that the shaded green areas cancel each
    other.  Representative values of the parameters are chosen; the
    unit of length is the sedimentation length $k_BT/(mg)$ and all
    energies are scaled with $k_BT$.}
  \label{FIG3}
\end{figure}

We exemplify the general result
\eqref{EQFextTotViaDensityProfileNoether} using the concrete example
of a thermal system under gravity, such that sedimentation-diffusion
equilibrium is reached. Recall that we consider systems at finite
temperature, where entropic effects compete with ordering generated by
the potential energy. We first omit the interparticle interactions,
and hence consider the classical monatomic ideal gas. We assume that
the external potential consists of a gravitational contribution,
$mgz$, where $g$ indicates the gravitational acceleration and $z$ is
the height variable. Furthermore due to the presence of a lower
container wall, there is a repulsive contribution, which we take to be
a harmonic potential with spring constant $\alpha$ acting ``inside''
the wall, i.e.\ at altitudes $z<0$. Hence the specific form of the
total external potential is
\begin{align}
  V_{\rm ext}(z) &= mgz + \frac{\alpha z^2}{2} \Theta(-z),
\end{align}
where $\Theta(\cdot)$ indicates the Heaviside (unit step) function,
which ensures that the parabolic potential only acts for $z<0$. There
is no need for the presence of an upper wall to close the system, as
gravity alone already ensures that $V_{\rm ext}\to\infty$ for
$z\to\infty$.  The magnitude of the external force field is obtained
as $-V'_{\rm ext}(z)=-mg-\alpha z\Theta(-z)$, see Fig.~\ref{FIG3} for
an illustration (blue line).

The density distribution of the isothermal ideal gas is given by the
generalized barometric law \cite{hansen2013},
\begin{align}
  \rho(z) &= \Lambda^{-3} {\rm e}^{-\beta(V_{\rm ext}(z)-\mu)},
  \label{EQbarometricLawNoether}
\end{align}
where $\Lambda$ is the thermal de Broglie wavelength which arises from
carrying out the momentum integrals in $\Tr_N$ (this is analytically
possible due to the simple kinetic energy part of the Boltzmann
factor). The chemical potential $\mu$ in
\eqref{EQbarometricLawNoether} is a constant that ensures the correct
normalization, $\int dz \rho(z)=N/A$, where $A$ is the lateral system
size (i.e.\ the area perpendicular to the $z$-direction).  That the
value of the chemical potential $\mu$ controls the number of particles
in the system is universal. However, the mathematical formulation in
the grand ensemble, where the particle number in the system can
fluctuate, is very different from the present canonical
treatment. (Some basics of the grand canonical description, as used in
Ref.~\cite{hermann2021noether}, are described below in
Sec.~\ref{SECdftpft}.)

The general expression for the total external force
\eqref{EQFextTotViaDensityProfileNoether} together with the specific
density profile \eqref{EQbarometricLawNoether} gives
\begin{align}
  \Fv_{\rm ext}^\tot &=
  -\frac{A \ev_z}{\Lambda^3}
  \int_{-\infty}^\infty dz 
      {\rm e}^{-\beta(V_{\rm ext}(z)-\mu)} V'_{\rm ext}(z)
   \label{EQidealGas1Noether}\\
  &= \frac{A\ev_z}{\Lambda^3\beta}
  \Big[{\rm e}^{-\beta(V_{\rm ext}(z)-\mu)}\Big]_{-\infty}^\infty
  \label{EQidealGas2Noether}
  \\
  &= 0, \label{EQidealGas3Noether}
\end{align}
where $\ev_z$ is the unit vector pointing into the positive
$z$-direction and the prime denotes differentiation with respect to
the argument, hence $\nabla V_{\rm ext}(\rv)=V_{\rm
  ext}'(z)\ev_z$. The integrand in \eqref{EQidealGas1Noether} is a
total differential, $d {\rm e}^{-\beta(V_{\rm ext}-\mu)}/dz$, which
upon integration gives \eqref{EQidealGas2Noether}; for
\eqref{EQidealGas3Noether} we have exploited that for $z\to\pm\infty$
the external potential $V_{\rm ext}\to\infty$, leading to vanishing
Boltzmann factor. We have hence shown explicitly the vanishing of the
total external force acting on a bounded ideal gas in thermal
equilibrium under gravity.  Figure \ref{FIG3} illustrates the density
profile $\rho(z)$ and the force density profile $-V_{\rm
  ext}'(z)\rho(z)$ for representative values of the parameters.

We briefly sketch the effect of interparticle interactions. On a
formal level, and returning to the general case of arbitrary form of
$V_{\rm ext}(\rv)$, the density profile is given by a modified form of
\eqref{EQbarometricLawNoether}, which reads
\begin{align}
  \rho(\rv) &= \Lambda^{-3}
  {\rm e}^{-\beta (V_{\rm ext}(\rv)-\mu) + c_1(\rv)},
  \label{EQbarometricLawNModifiedNoether}
\end{align}
where the so-called one-body direct correlation function
\cite{evans1979,hansen2013} $c_1(\rv)$ contains the effects of the
interparticle interactions. The total interparticle force density is
then given by
\begin{align}
  \Fv_{\rm int}^\tot &=
  k_BT \int d\rv \rho(\rv)\nabla c_1(\rv)=0.
  \label{EQFintTotViaDensityProfileNoether}
\end{align}
Here the vanishing of the total internal force can be viewed as a
consequence of Newtons' third law {\it actio equals reactio}; see
Ref.~\cite{hermann2021noether} for the derivation. Note the formal
similarity of the total external and total intrinsic force Noether sum
rules, cf.~\eqref{EQFextTotViaDensityProfileNoether} and
\eqref{EQFintTotViaDensityProfileNoether}.  The no-bootstrap theorem
\eqref{EQFintTotViaDensityProfileNoether} holds beyond equilibrium, as
shown in Ref.~\cite{hermann2021noether}, and it hence debunks any
swamp escape myths.

An alternative derivation of \eqref{EQFintTotViaDensityProfileNoether}
rests on the Noether invariance of the free energy, where the later is
constructed to be a functional of the density profile; we refer the
Reader to Ref.~\cite{hermann2021noether} for a description of these
considerations and comment briefly on the embedding into the
frameworks of classical density functional theory and power functional
theory below in Sec.~\ref{SECdftpft}.

\subsection{Relationship to correlation functions}
\label{SECrelationshipToCorrelationFunctions}
Global identities, such as the sum rules of vanishing external force
\eqref{EQFextTotViaDensityProfileNoether} and of vanishing internal
force \eqref{EQFintTotViaDensityProfileNoether}, can be used as a
starting point to obtain position-resolved identities. Functional
differentiation with respect to an appropriate field creates
dependence on position.  Integrating over these additional variables
(or ``root points'' \cite{hansen2013}) then yields novel global
identities. While we refer the reader to
Ref.~\cite{hermann2021noether} for this treatment, we wish to
demonstrate here the direct derivation of such global identities.

We stick to the canonical ensemble and as a specific case return to
our initial example of a thermal average, i.e.\ the global external
potential energy $U_{\rm ext}$, as equivalently expressed in compact
notation \eqref{EQaverageExternalEnergyNoether} or the explicitly
written out phase space integral
\eqref{EQaverageExternalEnergyNoetherExplicit}.  Let us shift! The
external energy in the new system is then given by
\begin{align}
  U_{\rm ext}(\eps) &= \Tr_N
  \frac{{\rm e}^{-\beta H(\eps)}}{Z_N(\eps)}
  \sum_{i=1}^N 
  V_{\rm ext}(\rv_i-\eps).
  \label{EQUextOfEpsilon}
\end{align}
We Taylor expand to first order,
\begin{align}
  U_{\rm ext}(\eps) &= U_{\rm ext} +
  \frac{\partial U_{\rm ext}(\eps)}{\partial \eps}
  \Big|_{\rm \eps=0}\cdot\eps.
  \label{EQUextTaylor}
\end{align}
Here the derivative of \eqref{EQUextOfEpsilon} can be calculated via
the product rule as
\begin{align}
  \frac{\partial U_{\rm ext}(\eps)}{\partial \eps}\Big|_{\eps=0}
  &= 
  \Tr\, \frac{\partial \Psi(\eps)}{\partial \eps}
  \sum_{i=1}^N V_{\rm ext}(\rv_i)\notag\\
  &\quad 
  -\Tr \, \Psi \sum_{i=1}^N \nabla_i V_{\rm ext}(\rv_i).
  \label{EQUextDerivativeIntermediate}
\end{align}
We can recognize the second term as the average external force, which
we have proven to vanish, cf.~\eqref{EQresultOfProofNoether}. The
first term in \eqref{EQUextDerivativeIntermediate} requires carrying
out the derivative of $\Psi(\eps)$ with respect to the displacement
$\eps$, which yields
\begin{align}
  \frac{\partial U_{\rm ext}(\eps)}{\partial \eps}\Big|_{\eps=0}
  &= 
  -\Tr \, \Psi
  \sum_{i=1}^N \beta V_{\rm ext}(\rv_i)
  \sum_{j=1}^N \nabla_j V_{\rm ext}(\rv_j).
  \label{EQexternalForceEnergyCorrelator}
\end{align}
Here an additional term, generated by the derivative, vanishes:
$-\beta U_{\rm ext}\Fv_{\rm ext}^\tot=0$, again due to
\eqref{EQresultOfProofNoether}.

Clearly \eqref{EQexternalForceEnergyCorrelator} is the correlator of
the global external potential energy and the global external
force. Using the by now familiar invariance argument, we argue that
the value of $U_{\rm ext}(\eps)$ is an invariant under the
displacement, and that hence the first order term in
\eqref{EQUextTaylor} needs to be zero. As $\eps$ is arbitrary, we
conclude
\begin{align}
  -\Tr \, \Psi
  \sum_{i=1}^N V_{\rm ext}(\rv_i)
  \sum_{j=1}^N \nabla_j V_{\rm ext}(\rv_j) &= 0,
  \label{EQexternalCrossCorrelatorZero}
\end{align}
where we have divided by $\beta$.  Hence the global external
potential, $\sum_i V_{\rm ext}(\rv_i)$, and the global external force,
$-\sum_j \nabla_j V_{\rm ext}(\rv_j)$, are uncorrelated with each
other. The sum rule \eqref{EQexternalCrossCorrelatorZero} is derived
in Ref.~\cite{hermann2021noether} via the route of integration over
free position variables (root points), cf.~(5) in
Ref.~\cite{hermann2021noether} for the order $n=2$ of the sum rule
hierarchy. An important distinction in the presentation though lies in
the choice of ensemble, which is an issue to which we turn
in the next subsection.

As a final comment, when applied to the above example of sedimentation
against a lower harmonic wall, \eqref{EQexternalCrossCorrelatorZero}
can be explicitly verified by carrying out the $z$-integral, which
yields $-A\int_{\infty}^\infty dz \rho(z) V_{\rm ext}(z) V'_{\rm
  ext}(z)=0$.

\subsection{Density functional and power functional}
\label{SECdftpft}
In all of the above, we have described the thermal system on the basis
of the canonical ensemble, as specified by the classical phase space,
the probability distribution \eqref{EQprobabilityDistributionNoether}
and the canonical partition sum \eqref{EQpartitionSumNoether}.  Hence
the system is coupled to a heat bath at temperature $T$, where the
value of $T$ determines the mean energy $E$ in the system, cf.\ the
form of $E$ as an expectation value
\eqref{EQaverageEnergyNoether}. The system is thermally open, and
hence energy fluctuations occur between system and bath.

Corresponding fluctuations in particle number $N$ can be implemented
in the grand canonical ensemble where the system is furthermore
coupled to a particle bath. The particle bath sets the value of the
chemical potential $\mu$, which then determines the average number of
particles $\bar N$ in the system. (This mechanism is analogous to the
relationship of $T$ and $E$ described above.) Although the grand
canonical formalism poses this additional level of abstraction, and
the bare formulae increase somewhat in complexity due to the average
over $N$, in typical theoretical developments this framework is
significantly more powerful and more straightforward to use. (There is
no need having to implement $N=\rm const$, which in practice can be
awkward.)  We briefly sketch the essentials of the grand ensemble as
they underlie Ref.~\cite{hermann2021noether}.

The grand canonical ensemble consists of the microstates given by
phase space points of $N$ particles, with $N$ being a non-negative
integer, which is treated as a random variable. The corresponding
probability distribution is
\begin{align}
  \Psi(\rv_1,\ldots,\rv_N,\pv_1,\ldots,\pv_N,N) &=
  \frac{{\rm e}^{-\beta(H-\mu N)}}{\Xi},
  \label{EQGCprobabilityDistributionNoether}
\end{align}
where the grand partition sum is given by
\begin{align}
  \Xi &= \Tr \, {\rm e}^{-\beta(H-\mu N)},
  \label{EQGCpartitionSumNoether}
\end{align}
with the grand canonical trace operation defined by
\begin{align}
  \Tr &= \sum_{N=0}^\infty \Tr_N
  \label{EQGCcanonicalTraceNoether}\\
  &= \sum_{N=0}^\infty \frac{1}{h^{3N}N!}
  \int d\rv_1\ldots d\rv_N d\pv_1\ldots d\pv_N,
  \label{EQGCcanonicalTraceExplicitNoether}
\end{align}
where we have obtained \eqref{EQGCcanonicalTraceExplicitNoether} by
using the explicit form \eqref{EQcanonicalTraceNoether} for the
canonical trace. The thermodynamic potential which is fundamental for
the grand ensemble is the grand potential (also referred to as the
grand canonical free energy) and it is given by
\begin{align}
  \Omega &= -k_BT \ln \Xi,
  \label{EQGComegaNoether}
\end{align}
with the grand partition sum $\Xi$ according to
\eqref{EQGCpartitionSumNoether}.  Note the strong formal analogy with
the corresponding canonical expressions for: the probability
distribution \eqref{EQprobabilityDistributionNoether} with
\eqref{EQGCprobabilityDistributionNoether}; the partition sum
\eqref{EQpartitionSumNoether}, i.e.\ $Z_N=\Tr_N {\rm e}^{-\beta H}$,
with \eqref{EQGCpartitionSumNoether}; the trace
\eqref{EQcanonicalTraceNoether} with
\eqref{EQGCcanonicalTraceNoether}; and the free energy
\eqref{EQfreeEnergyDefinitionNoether} with \eqref{EQGComegaNoether}.

Despite the system being open to particle exchange, Noether's
reasoning continues to hold~\cite{hermann2021noether}.  Briefly, the
grand potential is a functional of the external potential,
$\Omega[V_{\rm ext}]$ (we suppress the dependence on the thermodynamic
parameters $\mu,T$), and $\Omega[V_{\rm ext}]$ is invariant under
spatial displacements according to
\eqref{EQexternalPotentialDisplacementNoether}. As a consequence, the
sum rule of vanishing external force \eqref{EQforceZeroNoether}
emerges, expressed in the form \eqref{EQresultOfProofNoether} with
$\Tr_N$ replaced by $\Tr$, as is appropriate for the open system.

Why is the functional point of view important? In what we have
presented above it had played the role of adding abstraction and
re-deriving results that we could obtain via more elementary
arguments. The importance of the variational formulation stems from
two sources, one being that it provides a mechanism for the generation
of correlation functions via functional differentiation, in extension
of the generation of the density profile
via~\eqref{EQdensityProfileNoether}, see e.g.\ Refs.~\cite{hansen2013}
for a comprehensive account. The second point lies in the variational
principle itself which formulates the many-body problem in a way that
allows to systematically introduce approximations and make much
headway in identifying and studying physical mechanisms in complex,
coupled many-body problems. While giving a self-contained overview of
these concepts is beyond the scope of the present contribution (see
Ref.~\cite{schmidt2021pft} for a recent account), we wish to briefly
describe certain central points, to --hopefully-- provide motivation
for further study.

We hence sketch the two variational principles as they are relevant
for equilibrium (classical density functional theory) and for the
dynamics (power functional theory); these form the basis of
Ref.~\cite{hermann2021noether}. Classical density functional theory is
based on treating the density profile $\rho(\rv)$, rather than the
external potential $V_{\rm ext}(\rv)$, as the fundamental variational
field. The grand potential, when viewed as a density functional
\cite{evans1979,evans1992}, has the form
\begin{align}
  \Omega[\rho] &= 
  F[\rho] + \int d\rv \rho(\rv)(V_{\rm ext}(\rv)-\mu),
  \label{EQomegaDensityFunctionalNoether}
\end{align}
where $F[\rho]$ is the intrinsic Helmholtz free energy
functional. Crucially, $F[\rho]$ is independent of the external
potential, which features solely in the second term in
\eqref{EQomegaDensityFunctionalNoether}. Here $\rho(\rv)$ is
conceptually treated as a variable; its true form as the equilibrium
density profile is that which minimizes $\Omega[\rho]$ and for which
hence the functional derivative vanishes,
\begin{align}
  \frac{\delta \Omega[\rho]}{\delta \rho(\rv)} &= 0 \quad {\rm (min)}.
  \label{EQdftMinimizationPrincipleNoether}
\end{align}
Inserting the split form \eqref{EQomegaDensityFunctionalNoether} of
the grand potential into the minimization condition
\eqref{EQdftMinimizationPrincipleNoether} and using the splitting into
ideal gas and excess (over ideal gas) free energy contributions,
$F[\rho]=k_BT \int d\rv\rho(\rv)[\ln(\rho(\rv)\Lambda^3)-1] + F_{\rm
  exc}[\rho]$, yields upon exponentiating the modified barometric law
\eqref{EQbarometricLawNModifiedNoether}. Here the one-body direct
correlation function $c_1(\rv)$ is identified as the functional
derivative of the excess free energy functional,
i.e.\ $c_1(\rv)=-\beta \delta F_{\rm exc}[\rho]/\delta\rho(\rv)$.  As
the functional dependence on the density profile persists upon
building the derivative, i.e.\ in more explicit notation
$c_1(\rv,[\rho])$, equation \eqref{EQbarometricLawNModifiedNoether}
constitutes a self-consistency condition for the determination of the
equilibrium density profile; determining the solution thereof requires
to have an approximation for $F_{\rm exc}[\rho]$ and typically
involves numerical work.

Power functional theory generalizes the variational concept of working
on the level of one-body correlation functions to nonequilibrium. For
overdamped Brownian motion, as is a simple model for the description
for the temporal behaviour of mesoscopic particles that are suspended
in a liquid, the free power is a functional of both the time-dependent
density profile $\rho(\rv,t)$ and of the locally resolved current
distribution $\Jv(\rv,t)$, where $t$ indicates time. The power
functional has the form
\begin{align}
  R_t[\rho,\Jv] &= \dot F[\rho] + P_t[\rho,\Jv]
  \label{EQRtNoether}
  \\&\quad
  -\int d\rv (\Jv(\rv,t)\cdot \fv_{\rm ext}(\rv,t)
  -\rho(\rv,t) \dot V_{\rm ext}(\rv,t)),\notag
\end{align}
where $\dot F[\rho]$ is the time derivative of the intrinsic free
energy functional, $P_t[\rho,\Jv]$ consists of an ideal gas and a
superadiabatic part, where the latter arises from the internal
interactions in the nonequilibrium situation, $\fv_{\rm ext}(\rv,t)$
is a time-dependent external one-body force field, which in general
consists of a (conservative) gradient term $-\nabla V_{\rm
  ext}(\rv,t)$ and an additional rotational (non-gradient,
non-conservative) contribution, and $\dot V_{\rm ext}(\rv,t)$ is the
time derivative of the external potential. The density profile and the
current distribution are linked by the continuity equation, $\partial
\rho(\rv,t)=-\nabla\cdot\Jv(\rv,t)$, which is sharply resolved on the
microscopic scale.  The dynamic variational principle states that
$R_t[\rho,\Jv]$ is minimized, at time $t$, by the physically realized
current,
\begin{align}
  \frac{\delta R_t[\rho,\Jv]}{\delta\Jv(\rv,t)} &=
  0\quad {\rm (min)}.
  \label{EQpftMinimizationPrincipleNoether}
\end{align}
Inserting the splitting \eqref{EQRtNoether} of the total free power
into the minimization condition
\eqref{EQpftMinimizationPrincipleNoether} yields the 
formally exact force density relationship,
\begin{align}
  \gamma \Jv(\rv,t) &=
  -k_BT \nabla \rho(\rv,t)
  +\Fv_{\rm int}(\rv,t)
  +\rho(\rv,t)\fv_{\rm ext}(\rv,t),
  \label{EQforceDensityBalanceNoether}
\end{align}
where $\gamma$ is the friction constant of the overdamped motion, such
that the left hand side constitutes the (negative) friction force
density at position~$\rv$ and time~$t$. The right hand side of
\eqref{EQforceDensityBalanceNoether} consists of an ideal, an internal
and an external driving contribution, with $\Fv_{\rm int}(\rv,t)$
being the internal force density distribution, as it arises from the
effect of all interparticle interactions that act on a given particle
at position $\rv$ and time~$t$. The internal force density $\Fv_{\rm
  int}(\rv,t)$ consists of an adiabatic contribution, which follows
from the excess free energy functional via $\Fv_{\rm
  ad}(\rv,t)=-\rho(\rv,t)\nabla\delta F[\rho]/\delta\rho(\rv,t)$ and
an additional genuine nonequilibrium contribution, i.e.\ the
superadiabatic force density, $\Fv_{\rm sup}(\rv,t)$. Honoring its
functional dependence on the kinematic fields $\rho(\rv,t)$ and
$\Jv(\rv,t)$ forms the basis for much recent work in Nonequilibrium
Statistical Mechanics based on the power functional concept.

As a comment on terminology, we note that sometimes the term
Euler-Lagrange equation is applied generically to refer to the
vanishing of the first functional derivative of the given variational
problem, i.e.\ equation \eqref{EQdftMinimizationPrincipleNoether} for
the case of DFT and equation \eqref{EQpftMinimizationPrincipleNoether}
for PFT, which respectively turn into the explicit forms
\eqref{EQbarometricLawNModifiedNoether} and
\eqref{EQforceDensityBalanceNoether}. This terminology is different
from the also frequent use of referring specifically to the
Euler-Lagrange equations of motion of classical mechanics, as they
result from Hamilton's principle, i.e.\ the stationarity of the action
functional (see e.g.\ the appendix of Ref.~\cite{schmidt2021pft} for a
description of the functional methods involved).

\section{Conclusions}
\label{SECconclusions}

In conclusion, we have demonstrated on an elementary level how
fundamental symmetries in Statisical Mechanics lead to exact
statements (sum rules) about average forces when considering
translations. These considerations also apply to torques when
considering rotations \cite{hermann2021noether}. We have based our
presentation on the canonical ensemble, as is relevant in a variety of
contexts~\cite{gonzalez1997,gonzalez1998,delasheras2014fullCanonical,
  delasheras2016particleConservation,
  schindler2019,wittmann2021}. While the canonical ensemble avoids the
complexity of particle number fluctuations that occur grand
canonically, nevertheless an open system is retained with respect to
energy exchange with a heat bath.  As we have shown, treating such
fluctuating systems is well permissible on the basis of Noetherian
arguments. The arguably simplest Noether sum rule is that of vanishing
average total external force in thermal equilibrium. As an application
we have presented the case of a fluid confined inside of a container
and subject to the effect of gravity. While we have selected this
example for its relative simplicity, the influence of gravity on
mesoscopic soft matter is also a topic of relevance for studying
e.g.\ complex phase behaviour in colloidal mixtures; see
e.g.~Ref.~\cite{eckert2021} for recent work that addresses colloidal
liquid crystals.
Our derivations imply that the symmetry operation is applied to the
entire system. Here the system must be enclosed by an external
potential that represents confinement by e.g.\ walls. The shift then
applies also to these walls. In cases where system boundaries are open
(as can be suitable for a periodically repeated system like that shown
in Fig.~\ref{FIG2}), Noether's theorem remains applicable upon taking
account of additional boundary terms, see
Ref.~\cite{hermann2021noether} for a detailed discussion of such
treatment.

In the presented considerations, we have started on the basis of
arguably the most fundamental statistical mechanical object, i.e.\ the
partition sum, as it enters the elementary definition of the (here
canonical) free energy. Investigating invariance properties of further
statistical objects, such as the global external energy, is also
worthwhile, as then Noether's reasoning leads to the correlator
identity \eqref{EQexternalCrossCorrelatorZero} of vanishing
correlation between global external force and global external
potential. Investigating the outcome of invariance applied in this way
constitutes an interesting task for future work.

Statistical mechanical derivations often rely on very similar
reasoning; Ref.~\cite{hermann2021noether} gives an overview.  A
particularly insightful example is the work by Bryk \textit{et
  al.}\ on hard sphere fluids in contact with curved substrates
\cite{bryk2003}. These authors derive a contact sum rule of the hard
sphere fluid against a hard curved wall. Their argumentation rests on
the observation that the force that is necessary to move the wall by
an amount $\eps$ is balanced by the presence of the fluid. The authors
then succeed in relating this force to the value of the density
profile close to the wall. Closely related work was carried out for
the shape dependence of free energies \cite{konig2004}.  Further
studies that are related to Noether's Theorem were aimed at broken
symmetries \cite{baus1984} and emerging Goldstone
modes~\cite{walz2010,parry2016,parry2019}.

The general form of Noether's Theorem applies to variational calculus,
and Statistical Mechanics falls well into this realm. We have spelled
out the connections explicitly, such as the canonical free energy
being viewed as a functional of the external potential
\cite{hansen2013}. Notably only elementary statistical objects such as
the partition sum are required. We have also described two more
advanced variational theories. Classical density functional theory
\cite{evans1979,evans1992,hansen2013} allows to view the grand
potential as a functional of the one-body density distribution. A
formally exact minimization principle then reformulates the physics of
system in thermal (and chemical) equilibrium.  The dynamic variational
principle of power functional theory
\cite{schmidt2013pft,schmidt2021pft} consists of instantaneous
minimization with respect to the time- and position-resolved current
distribution. Together with the continuity equation, a formally closed
one-body reformulation of the dynamics of the underlying many-body
system is achieved.

Both density functional theory and power functional theory can be
viewed as systematic approaches to coarse-graining the many-body
problem to the level of one-body correlation functions. In the static
case, the correlation functions hence depend on position alone, in the
dynamics case the dependence is on position and on time. Crucially, a
microscopically sharp description is formally retained, which is
important for the description of correlations on the particle
(i.e.\ molecular or colloidal) level. One of the most important
features of these theories is the identification of a universal
intrinsic functional that contains the coupled effects of the
interparticle interactions, but is independent of the external forces
that act on the system.

A wealth of productive research have been devoted to constructing
powerful approximations for free energy functionals for specific model
systems. In the context of liquids the important case of the hard
sphere fluid is treated with excellent accuracy within Rosenfeld's
fundamental measure theory \cite{rosenfeld1989,roth2010review}, see
e.g.\ Ref.~\cite{davidchack2016} for a quantitative assessment of the
quality of theoretical density profiles against simulation
data. Notable recent progress to incorporate short-ranged attraction
into density functional theory is due to Tschopp, Brader and their
coworkers \cite{tschopp2020,tschopp2021}, who systematically addressed
and exploited two-body correlations.

Despite power functional theory \cite{schmidt2013pft,schmidt2021pft}
being significantly younger than density functional theory, its
usefulness has been amply demonstrated, both for formal work as well
as for practical solution of physical problems and the discovery of
novel fundamental mechanisms. The reformulation on the basis of the
velocity gradient \cite{delasheras2018velocityGradient}, instead of
the current distribution, allowed to identify and to study structural
forces \cite{stuhlmueller2018prl,delasheras2020prl} in driven systems
that are governed by overdamped Brownian dynamics. The splitting of
the total internal force field into flow and structural contributions
is fundamental to understanding the emerging effects in
microscopically inhomogeneous flows \cite{delasheras2020prl}.  Active
Brownian particles, as a model for self-propelled colloids (see
e.g.~\cite{farage2015,paliwal2017activeLJ,paliwal2018}), are well
suited for the application of power functional theory. The general
framework \cite{krinninger2016prl,krinninger2019jcp} for active
systems was shown to physically explain and quantitatively predict the
motility-induced phase separation that occurs in such systems at high
enough levels of driving
\cite{hermann2019pre,hermann2019prl,hermann2020longActive}. Interfacial
properties such as polarization \cite{hermann2020polarization} and
surface tension \cite{hermann2019prl} were systematically studied.

The dynamical sum rules for forces and correlation functions presented
in Ref.~\cite{hermann2021noether} offer great potential for systematic
progress in the description of complex temporal behaviour, including
memory \cite{treffenstaedt2020shear,treffenstaedt2021dtpl}.  The
nonequilibrium rules play a similar role than fundamental equilibrium
sum rules such e.g.\ the Lovett-Mou-Buff-Wertheim equation
\cite{lovett1976,wertheim1976}. The section on ``Methods'' in
Ref.~\cite{hermann2021noether} gives a detailed description of the
relationship of the equilibrium Noether sum rules to to such classical
results from the liquid state literature.  Together with the
nonequilibrium Ornstein-Zernike relations
\cite{brader2013noz,brader2014noz} the dynamical sum rules provide
fertile ground for making progress in nonequilibrium many-body
physics; see also the recent study of the relevance of invariance in
inhomogeneous dense liquids \cite{tschopp2022forceDFT} and of the role
of fluctuations when going to effects that are higher than linear in
the displacement \cite{hermann2022invariance}. Hence the fundamental
character of Emmy Noether's work will surely continue to prove its
worth in the future.

\vspace{5mm}
\acknowledgments We thank Daniel de las Heras, Roland Roth, and Bob
Evans for useful comments and discussions. This work is supported by
the German Research Foundation (DFG) via project number 436306241.


\begin{thebibliography}{31}


\bibitem{noether1918}
  Invariante Variationsprobleme.
  E. Noether,
  \href{https://gdz.sub.uni-goettingen.de/download/pdf/PPN252457811_1918/LOG_0022.pdf}
       {Nachr. d. K\"onig. Gesellsch. d. Wiss. zu G\"ottingen, 
  Math.-Phys. Klasse, 235 (1918).}
%
  English translation by M. A. Tavel: Invariant variation
  problems. Transp. Theo. Stat.  Phys. {\bf 1}, 186 (1971); for a
  version in modern typesetting see: Frank Y. Wang,
  \href{http://arxiv.org/abs/physics/0503066v3}{arXiv:physics/0503066v3}
  (2018).

\bibitem{neuenschwander2011} 
  For a description of many insightful and
  pedagogical examples and applications, see: D. E. Neuenschwander,
  {\it Emmy Noether's Wonderful Theorem} (Johns Hopkins University
  Press, Baltimore, 2011).

\bibitem{byers1998} 
  N. Byers,
  ``E.\ Noether's discovery of the deep connection between
  symmetries and conservation laws,''
  \href{https://arxiv.org/abs/physics/9807044}
       {arXiv:physics/9807044 (1998)}.


\bibitem{lezcano2018stochastic}
  A. G. Lezcano and A. C. M. de Oca, 
  ``A Stochastic Version of the Noether Theorem,'' 
  \href{https://doi.org/10.1007/s10701-018-0174-z}
       {Found. Phys. \textbf{48}, 726 (2018).}

\bibitem{baez2013markov}
  J. C. Baez and B. Fong, 
  ``A Noether theorem for Markov processes,'' 
  \href{http://dx.doi.org/10.1063/1.4773921}
       {J. Math. Phys. \textbf{54}, 013301 (2013).}

\bibitem{marvian2014quantum}
  I. Marvian and R. W. Spekkens, 
``Extending Noether’s theorem by quantifying the asymmetry of quantum states,''
  \href{https://doi.org/10.1038/ncomms4821}
       {Nat. Commun. \textbf{5}, 3821 (2014).}

\bibitem{sasa2016}
   S. Sasa and Y. Yokokura,
   ``Thermodynamic Entropy as a Noether Invariant,''
   \href{http://dx.doi.org/10.1103/PhysRevLett.116.140601}
   {Phys. Rev. Lett. {\bf 116}, 140601 (2016).}
   
\bibitem{sasa2019} 
S. Sasa, S. Sugiura, and Y. Yokokura, 
``Thermodynamical path integral and emergent symmetry,'' 
\href{https://doi.org/10.1103/PhysRevE.99.022109}
{Phys. Rev. E \textbf{99}, 022109 (2019).}

\bibitem{sasa2020}
Y. Minami and S. Sasa, 
``Thermodynamic entropy as a Noether invariant in a Langevin equation,'' 
\href{https://doi.org/10.1088/1742-5468/ab5b8b}
{J. Stat. Mech. \textbf{2020}, 013213 (2020).}


\bibitem{mermin1965}
   N. D. Mermin, 
   ``Thermal properties of the inhomogeneous electron gas,''
   \href{https://doi.org/10.1103/PhysRev.137.A1441}
        {Phys. Rev. {\bf 137}, A1441 (1965).}

\bibitem{evans1979}
  R. Evans,
  ``The nature of the liquid-vapour interface and other topics in the statistical
  mechanics of non-uniform, classical fluids,''
  \href{https://doi.org/10.1080/00018737900101365}
       {Adv. Phys. {\bf 28},  143 (1979).}
       
\bibitem{evans1992}
  R. Evans,
  ``Density functionals in the theory nonuniform fluids,''
  in {\it Fundamentals of Inhomogeneous Fluids},
  edited by D. Henderson (Dekker, New York, 1992).

\bibitem{hansen2013}
  J. P. Hansen and I. R. McDonald, 
  {\it Theory of Simple Liquids}, 4th ed. (Academic Press, London, 2013).

      
\bibitem{hermann2021noether}
  S. Hermann and M. Schmidt, 
  ``Noether's Theorem in Statistical Mechanics,''
  \href{https://doi.org/10.1038/s42005-021-00669-2}
       {Commun. Phys. {\bf 4}, 176 (2021).}
See also:  \href{https://www.mschmidt.uni-bayreuth.de/noether.html}
   {www.mschmidt.uni-bayreuth.de/noether.html}

\bibitem{tarazona1983}
    P. Tarazona and R. Evans, 
    ``On the validity of certain integro-differential equations
  for the density-orientation profile of molecular fluid interfaces,''
  \href{https://doi.org/10.1016/0009-2614(83)80007-4}
       {Chem. Phys. Lett. {\bf 97}, 279 (1983).}
   

\bibitem{evans2016specialIssue}
  R. Evans, M. Oettel, R. Roth, and G. Kahl, 
  ``New developments in classical density functional theory,''
  \href{https://doi.org/10.1088/0953-8984/28/24/240401}
  {J. Phys.: Condens. Matter {\bf 28}, 240401 (2016).}

\bibitem{rosenfeld1989}
  Y. Rosenfeld,
  ``Free-energy model for the inhomogeneous hard-sphere fluid mixture and
  density-functional theory of freezing,''
  \href{https://doi.org/10.1103/PhysRevLett.63.980}
       {Phys. Rev. Lett. {\bf 63}, 980 (1989).}

\bibitem{roth2010review}
  R. Roth,  
  ``Fundamental measure theory for hard-sphere mixtures: a review,''
  \href{https://doi.org/10.1088/0953-8984/22/6/063102}
       {J. Phys.: Condens. Matter {\bf 22}, 063102 (2010)}

\bibitem{lin2021}
  S.-C. Lin, M. Oettel, J. M. H{\"a}ring, R. Haussmann, M. Fuchs, and G. Kahl, 
  ``The direct correlation function of a crystalline solid,''
  \href{https://doi.org/10.1103/PhysRevLett.127.085501}
       {Phys. Rev. Lett. {\bf 127}, 085501 (2021).}

\bibitem{levesque2012jcp}
  M. Levesque, R. Vuilleumier, and D. Borgis,  
  ``Scalar fundamental measure theory for hard spheres in three
  dimensions: Application to hydrophobic solvation,''
  \href{https://doi.org/10.1063/1.4734009}
  {J. Chem. Phys. {\bf 137}, 034115 (2012).}

\bibitem{jeanmairet2013jcp}
  G. Jeanmairet, M. Levesque, and D. Borgis,  
  ``Molecular density functional theory of water describing 
  hydrophobicity at short and long length scales,'' 
  \href{https://doi.org/10.1063/1.4824737}
  {J. Chem. Phys. {\bf 139}, 154101  (2013).}


\bibitem{evans2019pnas}
  R. Evans, M. C. Stewart, and N. B. Wilding,  
  ``A unified description of hydrophilic and superhydrophobic surfaces in 
  terms of the wetting and drying transitions of liquids,''
  \href{https://doi.org/10.1073/pnas.1913587116}
  {Proc. Nat. Acad. Sci. {\bf 116}, 23901 (2019).}


\bibitem{evans2015prl}
  R. Evans and N. B. Wilding,  
  ``Quantifying density fluctuations in water at a hydrophobic surface:
  evidence for critical drying,''
  \href{http://doi.org/10.1103/PhysRevLett.115.016103}
  {Phys. Rev. Lett. {\bf 115}, 016103 (2015).}

\bibitem{evans2016prl}
  R. Evans, M. C. Stewart, and N. B. Wilding, 
  ``Critical drying of liquids,''
  \href{http://doi.org/10.1103/PhysRevLett.117.176102}
  {Phys. Rev. Lett. {\bf 117}, 176102 (2016).}

\bibitem{chacko2017}
  B. Chacko, R. Evans, and A. J. Archer, 
  ``Solvent fluctuations around solvophobic, solvophilic, and patchy
  nanostructures and the accompanying solvent mediated interactions,''
  \href{https://doi.org/10.1063/1.4978352}
  {J. Chem. Phys. {\bf 146}, 124703 (2017).}

\bibitem{martinjimenez2017natCom} 
  D. Martin-Jimenez, E. Chac\'on, P. Tarazona, and R. Garcia, 
  ``Atomically resolved three-dimensional structures of
  electrolyte aqueous solutions near a solid surface,''
  \href{https://doi.org/10.1038/ncomms12164}
  {Nat. Commun. {\bf 7}, 12164 (2016).}

\bibitem{hernandez-munoz2019}
  J. Hern\'andez-Mu\~noz, E. Chac\'on, and P. Tarazona, 
  ``Density functional analysis of atomic force microscopy in a dense fluid,''
   \href{https://doi.org/10.1063/1.5110366}
   {J. Chem. Phys. {\bf 151}, 034701 (2019).}

\bibitem{muscatello2017}
  J. Muscatello, E. Chac\'on, P. Tarazona, and F. Bresme, 
  ``Deconstructing temperature gradients across fluid
  interfaces: the structural origin of the thermal resistance of
  liquid-vapor interfaces,''
  \href{https://doi.org/10.1103/PhysRevLett.119.045901}
  {Phys. Rev. Lett. {\bf 119}, 045901 (2017).}
  

\bibitem{eckert2020auxiliaryFields}
  T. Eckert, N.~C.~X. Stuhlm\"uller, F. Samm\"uller, and M. Schmidt,
  ``Fluctuation profiles in inhomogeneous fluids,''
  \href{https://doi.org/10.1103/PhysRevLett.125.268004}
       {Phys. Rev. Lett. {\bf 125}, 268004 (2020).}

\bibitem{schmidt2013pft}
  M. Schmidt and J. M. Brader,   
  ``Power functional theory for Brownian dynamics,''
  \href{https://doi.org/10.1063/1.4807586}
       {J. Chem. Phys. {\bf 138}, 214101 (2013).}

\bibitem{schmidt2021pft}
  M. Schmidt,
  ``Power functional theory for many-body dynamics,''
  Rev. Mod. Phys. (to appear);
  \href{https://arxiv.org/abs/2111.00432}
  {arxiv:2111.00432.}


\bibitem{fortini2014prl}
  A. Fortini, D. de las Heras, J. M. Brader, and M. Schmidt, 
  ``Superadiabatic forces in Brownian many-body dynamics,''
  \href{https://doi.org/10.1103/PhysRevLett.113.167801}
       {Phys. Rev. Lett. {\bf 113}, 167801 (2014).}

\bibitem{krinninger2016prl}
  P. Krinninger, M. Schmidt, and J. M. Brader, 
  ``Nonequilibrium phase behaviour from minimization of free power dissipation,'' 
  \href{https://doi.org/10.1103/PhysRevLett.117.208003}
  {Phys. Rev. Lett. {\bf 117}, 208003 (2016).}

\bibitem{krinninger2019jcp}
  P. Krinninger  and M. Schmidt, 
  ``Power functional theory for active Brownian particles:
  general formulation and power sum rules,''
  \href{https://doi.org/10.1063/1.5061764}
       {J. Chem. Phys. {\bf 150}, 074112 (2019).}

\bibitem{hermann2019prl}
  S. Hermann, D. de las Heras, and M. Schmidt, 
  ``Non-negative interfacial tension in phase-separated active Brownian particles,''
  \href{https://doi.org/10.1103/PhysRevLett.123.268002}
       {Phys. Rev. Lett. {\bf 123}, 268002 (2019).}

\bibitem{hermann2019pre}
  S. Hermann, P. Krinninger, D. de las Heras, and M. Schmidt, 
  ``Phase coexistence of active Brownian particles,''
  \href{https://doi.org/10.1103/PhysRevE.100.052604}
       {Phys. Rev. E {\bf 100}, 052604 (2019).}

\bibitem{hermann2020longActive}
  S. Hermann D. de las Heras, and M. Schmidt, 
  ``Phase separation of active Brownian particles in two dimensions: 
  Anything for a quiet life,''
  Mol. Phys.
  \href{https://doi.org/10.1080/00268976.2021.1902585}{e1902585}; see also:
  \href{https://arxiv.org/abs/2103.03585}{arxiv:2103.03585} (2020).

\bibitem{hermann2020polarization}
  S. Hermann and M. Schmidt,  
  ``Active interface polarization as a state function,''
  \href{https://doi.org/10.1103/PhysRevResearch.2.022003}
       {Phys. Rev. Research {\bf 2}, 022003(R) (2020).}

\bibitem{delasheras2018velocityGradient}
 D. de las Heras and M. Schmidt,  
  ``Velocity gradient power functional for Brownian dynamics,'' 
 \href{https://doi.org/10.1103/PhysRevLett.120.028001}
      {Phys. Rev. Lett. {\bf 120}, 028001 (2018).}

\bibitem{stuhlmueller2018prl}
 N. C. X. Stuhlm\"uller, T. Eckert, D. de las Heras, and M. Schmidt,  
  ``Structural nonequilibrium forces in driven colloidal systems,'' 
 \href{https://doi.org/10.1103/PhysRevLett.121.098002}
      {Phys. Rev. Lett. {\bf 121}, 098002 (2018).}

\bibitem{delasheras2020prl}
  D. de las Heras and M. Schmidt,
  ``Flow and structure in nonequilibrium Brownian many-body systems,''
  \href{https://doi.org/10.1103/PhysRevLett.125.018001}
       {Phys. Rev. Lett. {\bf 125}, 018001 (2020).}

\bibitem{brader2013noz}
  J. M. Brader and M. Schmidt, 
  ``Nonequilibrium Ornstein-Zernike relation for Brownian many-body dynamics,''
  \href{https://doi.org/10.1063/1.4820399}
       {J. Chem. Phys. {\bf 139}, 104108 (2013).}

\bibitem{brader2014noz}
  J. M. Brader and M. Schmidt, 
  ``Dynamic correlations in Brownian many-body systems,''
  \href{https://doi.org/10.1063/1.4861041}
       {J. Chem. Phys. {\bf 140}, 034104 (2014).}

\bibitem{treffenstaedt2020shear}
  L. L. Treffenst\"adt and M. Schmidt,
  ``Memory-induced motion reversal in Brownian liquids,''
  \href{https://doi.org/10.1039/C9SM02005E}
       {Soft Matter {\bf 16}, 1518 (2020).}

\bibitem{treffenstaedt2021dtpl}
  L. L. Treffenst\"adt and M. Schmidt,
  ``Universality in driven and equilibrium hard sphere liquid dynamics,''
  \href{https://doi.org/10.1103/PhysRevLett.126.058002}
       {Phys. Rev. Lett. {\bf 126}, 058002 (2021).}







\bibitem{farage2015}
  T. F. F. Farage, P. Krinninger, and J. M. Brader,   
  ``Effective interactions in active Brownian suspensions,''
  \href{https://doi.org/10.1103/PhysRevE.91.042310}
       {Phys. Rev. E {\bf 91}, 042310 (2015).}

\bibitem{paliwal2018}
  S. Paliwal, J. Rodenburg, R. van Roij, and M. Dijkstra, 
  ``Chemical potential in active systems: predicting phase equilibrium from 
  bulk equations of state?''
  \href{https://doi.org/10.1088/1367-2630/aa9b4d}
       {New J. Phys. {\bf 20}, 015003 (2018).}

\bibitem{paliwal2017activeLJ}
  S. Paliwal, V. Prymidis, L. Filion, and M. Dijkstra, 
  ``Non-equilibrium surface tension of the vapour-liquid interface of 
  active Lennard- Jones particles,''
  \href{https://doi.org/10.1063/1.4989764}
       {J. Chem. Phys. {\bf 147}, 084902 (2017).}
  
\bibitem{brady2014}
  S. C. Takatori, W. Yan, and J. F. Brady, 
  ``Swim pressure: stress generation in active matter,'' 
  \href{https://doi.org/10.1103/PhysRevLett.113.028103}
       {Phys. Rev. Lett. \textbf{113}, 028103 (2014).}

\bibitem{hermann2018activeSedimentation}
  S. Hermann and M. Schmidt, 
  ``Active ideal sedimentation: exact two-dimensional steady states,'' 
  \href{https://doi.org/10.1039/C7SM02515G}
       {Soft Matter \textbf{14}, 1614 (2018).}

\bibitem{soeker2021}
  N. A. Söker, S. Auschra, V. Holubec, K. Kroy, and F. Cichos,
  ``How Activity Landscapes Polarize Microswimmers without Alignment Forces,''
  \href{https://doi.org/10.1103/PhysRevLett.126.228001}
  {Phys. Rev. Lett. \textbf{126}, 228001 (2021).}

\bibitem{auschra2021}
  S. Auschra, V. Holubec, N. A. Söker, F. Cichos, and K. Kroy, 
  ``Polarization-density patterns of active particles in motility gradients,''
  \href{https://doi.org/10.1103/PhysRevE.103.062601}
  {Phys. Rev. E \textbf{103}, 062601 (2021).}



\bibitem{gonzalez1997}
  A. Gonzalez, J. A. White, F. L. Roman, S. Velasco, R. Evans,
  ``Density functional theory for small systems: Hard spheres in a closed 
  spherical cavity,''
  \href{https://doi.org/10.1103/PhysRevLett.79.2466}
  {Phys. Rev. Lett. {\bf 79}, 2466 (1997).}

\bibitem{gonzalez1998}
  A. Gonzalez, J. A. White, F. L. Roman, R. Evans,
  ``How the structure of a confined fluid depends on the ensemble: 
  Hard spheres in a spherical cavity,''
  \href{https://doi.org/10.1063/1.476961}
       {J. Chem. Phys. {\bf 109}, 3637 (1998).}


\bibitem{delasheras2014fullCanonical}
  D. de las Heras and M. Schmidt, 
  ``Full canonical information from grand potential density functional theory,''
  \href{https://doi.org/10.1103/PhysRevLett.113.238304}
       {Phys. Rev. Lett. {\bf 113}, 238304 (2014).}

\bibitem{delasheras2016particleConservation}
  D. de las Heras, J. M. Brader, A. Fortini, and M.~Schmidt,
  ``Particle conservation in dynamical density functional theory,''
  \href{https://doi.org/10.1088/0953-8984/28/24/244024}
       {J. Phys.: Condens. Matter {\bf 28}, 244024 (2016).}

\bibitem{schindler2019}
  T. Schindler, R. Wittmann, and J. M. Brader,
  ``Particle-conserving dynamics on the single-particle level,''
  \href{https://doi.org/10.1103/PhysRevE.99.012605}
       {Phys. Rev. E {\bf 99}, 012605 (2019).}

\bibitem{wittmann2021}
  R. Wittmann, H. L{\"o}wen, and J. M. Brader,
  ``Order-preserving dynamics in one dimension -- single-file
  diffusion and caging from the perspective of 
  dynamical density functional theory,''
  \href{https://doi.org/10.1080/00268976.2020.1867250}
       {Mol. Phys. e1867250 (2021).}


\bibitem{zwanzig2001}
  R. Zwanzig,
  {\it Nonequilibrium Statistical Mechanics}
  (Oxford University Press, Oxford, 2001); see Chapter 10.

\bibitem{eckert2021}
  T. Eckert, M. Schmidt, and D. de las Heras, 
  ``Gravity-induced phase phenomena in plate-rod colloidal mixtures,''
  \href{https://doi.org/10.1038/s42005-021-00706-0}
       {Commun. Phys. \textbf{4}, 202 (2021).}


\bibitem{davidchack2016}
  R. L. Davidchack, B. B. Laird, and R. Roth,
  ``Hard spheres at a planar hard wall: Simulations and
  density functional theory,''
  \href{https://doi.org/10.5488/CMP.19.23001}
       {Condens. Matt. Phys. {\bf 19}, 23001 (2016).}

\bibitem{tschopp2020}
  S. M. Tschopp, H. D. Vuijk, A. Sharma, and J. M. Brader, 
  ``Mean-field theory of inhomogeneous fluids,'' 
  \href{https://doi.org/10.1103/PhysRevE.102.042140}
       {Phys. Rev. E {\bf 102}, 042140 (2020).}

\bibitem{tschopp2021}
  S. M. Tschopp and J. M. Brader,
  ``Fundamental measure theory of inhomogeneous two-body correlation functions,'' 
  \href{https://doi.org/10.1103/PhysRevE.103.042103}
       {Phys. Rev. E \textbf{103}, 042103 (2021).}

\bibitem{bryk2003}
  P. Bryk, R. Roth, K. R. Mecke, and S. Dietrich,
  ``Hard-sphere fluids in contact with curved substrates,'' 
  \href{https://doi.org/10.1103/PhysRevE.68.031602}
       {Phys. Rev. E \textbf{68}, 031602 (2003).}

\bibitem{konig2004}
  P.-M. K\"onig, R. Roth, and K. R. Mecke,
  ``Morphological Thermodynamics of Fluids: Shape Dependence of Free Energies,''
  \href{https://doi.org/10.1103/PhysRevLett.93.160601}
  {Phys. Rev. Lett. {\bf 93}, 160601 (2004).}

\bibitem{baus1984}
  M. Baus, 
  ``Broken symmetry and invariance properties of classical fluids,''
  \href{https://doi.org/10.1080/00268978400100161}
  {Mol. Phys. {\bf 51}, 211 (1984).}

\bibitem{walz2010}
  C. Walz and M. Fuchs, 
  ``Displacement field and elastic constants in nonideal crystals,''
  \href{http://dx.doi.org/10.1103/PhysRevB.81.134110}
  {Phys. Rev. B {\bf 81}, 134110 (2010).}

\bibitem{parry2016}
  A. O. Parry, C. Rasc\'on, and R. Evans,
  ``The local structure factor near an interface;
    beyond extended capillary-wave models,''
  \href{http://dx.doi.org/10.1088/0953-8984/28/24/244013}
  {J. Phys.: Condens. Matter {\bf 28}, 244013 (2016).}

\bibitem{parry2019}
  A. O. Parry and C. Rasc\'on.
  ``The Goldstone mode and resonances in the fluid interfacial region,''
  \href{https://doi.org/10.1038/s41567-018-0361-z}
  {Nature Physics {\bf 15}, 287 (2019).}

\bibitem{lovett1976}
  R. A. Lovett, C. Y. Mou, and F. P. Buff,
  ``The structure of the liquid–vapor interface,''
  \href{https://doi.org/10.1063/1.433110}
  {J. Chem. Phys. {\bf 65}, 570 (1976).}

\bibitem{wertheim1976}
  M. S. Wertheim,
  ``Correlations in the liquid–vapor interface,''
  \href{https://doi.org/10.1063/1.433352}
  {J. Chem. Phys. {\bf 65}, 2377 (1976).}

\bibitem{tschopp2022forceDFT}
  S. M. Tschopp, F. Samm\"uller, S. Hermann, M. Schmidt and J. M. Brader,
  ``Force density functional theory in- and out-of-equilibrium''
  (unpublished).

\bibitem{hermann2022invariance}
  S. Hermann and M. Schmidt,
  ``Variance of fluctuations from Noether invariance'' 
  (unpublished).

\end{thebibliography}
\end{document}